\documentclass[10pt,twocolumn,twoside,journal]{IEEEtran}

\usepackage{subfigure}
\usepackage{multirow}
\usepackage{graphicx}
\usepackage{amsmath}
\usepackage{color}
\usepackage{epsfig}
\usepackage{amsfonts}
\usepackage{amssymb}
\usepackage{amsthm}
\usepackage{algorithm2e}
\usepackage{algorithmic}

\newcommand{\argmax}{\mathop{\text{argmax}}}
\newcommand{\argmin}{\mathop{\text{argmin}}}

\newcommand{\vh}{{\bf h}}
\newcommand{\vw}{{\bf w}}
\newcommand{\vz}{{\bf z}}
\newcommand{\vx}{{\bf x}}
\newcommand{\vy}{{\bf y}}
\newcommand{\vs}{{\bf s}}
\newcommand{\vn}{{\bf n}}
\newcommand{\vmu}{{\boldsymbol \mu}}
\newcommand{\vp}{{\bf p}}
\newcommand{\vze}{\widehat{\bf z}}
\newcommand{\vv}{{\bf v}}
\newcommand{\vg}{{\bf g}}

\newcommand{\msig}{{\bf \Sigma}}
\newcommand{\mh}{{\bf H}}
\newcommand{\mi}{{\bf I}}

\newcommand{\sap}{{\cal S}}

\newcommand{\gsm}{{\mathbb S}_{n_t,{\mathbb A}}^{n_{rf}}}
\newcommand{\gsmk}{{\mathbb G}}
\newcommand{\sa}{{\mathbb A}}
\newcommand{\sj}{{\mathbb J}}
\newcommand{\sL}{{\mathbb L}}
\newcommand{\mj}{{\bf J}}
\newcommand{\mje}{\widehat{\bf J}}
\newcommand{\mx}{{\bf X}}
\newcommand{\mw}{{\bf W}}
\newcommand{\my}{{\bf Y}}

\newcommand{\E}{{\mathbb E}}

\newcommand{\Define}{\triangleq}

\begin{document}

\title{{\Huge 
Generalized Spatial Modulation in Large-Scale Multiuser MIMO Systems
}}
\author{
T. Lakshmi Narasimhan, P. Raviteja, and A. Chockalingam \\ 

Department of ECE, Indian Institute of Science, Bangalore-560012, India 
}

\maketitle

\begin{abstract}
Generalized spatial modulation (GSM) uses $n_t$ transmit antenna elements 
but fewer transmit radio frequency (RF) chains, $n_{rf}$. Spatial modulation 
(SM) and spatial multiplexing are special cases of GSM with $n_{rf}=1$ and 
$n_{rf}=n_t$, respectively. In GSM, in addition to conveying information 
bits through $n_{rf}$ conventional modulation symbols (for example, QAM), 
the indices of the $n_{rf}$ active transmit antennas also convey information 
bits. In this paper, we investigate {\em GSM for large-scale multiuser 
MIMO communications on the uplink}. Our contributions in this paper 
include: ($i$) an average bit error probability (ABEP) analysis for 
maximum-likelihood detection in multiuser GSM-MIMO on the uplink, where 
we derive an upper bound on the ABEP, and ($ii$) low-complexity algorithms 
for GSM-MIMO signal detection and channel estimation at the base station 
receiver based on message passing. The analytical upper bounds on the ABEP 
are found to be tight at moderate to high signal-to-noise
ratios (SNR). The proposed receiver algorithms are found to scale very 
well in complexity while achieving near-optimal performance in large 
dimensions. Simulation results show that, for the same spectral 
efficiency, multiuser GSM-MIMO can outperform multiuser SM-MIMO as well 
as conventional multiuser MIMO, by about 2 to 9 dB at a bit error rate 
of $10^{-3}$. Such SNR gains in GSM-MIMO compared to SM-MIMO and 
conventional MIMO can be attributed to the fact that, because of a 
larger number of spatial index bits, GSM-MIMO can use a lower-order 
QAM alphabet which is more power efficient. 
\end{abstract}


\vspace{-2mm}
\section{Introduction}
\label{sec1}
Large-scale MIMO systems with tens to hundreds of antennas are getting
increased research attention \cite{book}-\cite{scale2}. Because of its
advantages of very high spectral efficiencies/sum rates, 
increased reliability, and power efficiency, large-scale MIMO technology 
is being considered as a potential technology for next generation (example, 
5G) wireless systems \cite{5g}. The following two characteristics are 
typical in conventional MIMO systems: $(i)$ there will be one transmit 
radio frequency (RF) chain for each transmit antenna (i.e., if $n_t$ is 
the number of transmit antennas, then the number of transmit RF chains, 
$n_{rf}$, will also be $n_t$), and $(ii)$ information bits are carried 
only on the modulation symbols (example, QAM). Conventional multiuser MIMO 
systems with a large number (tens to hundreds) of antennas at the base 
station (BS) are referred to as `massive MIMO' systems in the recent 
literature \cite{scale2},\cite{5g}. Key technological issues that need 
to be addressed in practical realization of large-scale MIMO systems 
include design and placement of compact antennas, multiple RF chains, 
and large-dimension transmit/receive signal processing techniques and 
algorithms \cite{book}-\cite{scale2}. 

Spatial modulation (SM), an attractive modulation scheme for multi-antenna 
communications \cite{SM_commag},\cite{lajos}, can alleviate the requirement
of multiple transmit RF chains in MIMO systems. In SM, the transmitter has 
multiple transmit antennas but only one transmit RF chain. In a given channel
use, only one of the $n_t$ transmit antennas will be activated, and the 
remaining $n_t-1$ antennas remain silent. On the active transmit antenna, a 
symbol from a conventional modulation alphabet ${\mathbb A}$ is transmitted. 
In addition to information bits conveyed through the modulation symbol from 
${\mathbb A}$, the index of the active transmit antenna also conveys 
information bits. Therefore, the number of bits conveyed in one channel 
use in SM is $\lfloor\log_2n_t\rfloor+\lfloor\log_2 |{\mathbb A}|\rfloor$. 
Space shift keying (SSK) is a special case of SM. Like in SM, 
in SSK also only one antenna among $n_t$ antennas is activated in a given 
channel use. On the activated antenna, instead of sending a symbol from a 
conventional alphabet as is done in SM, a constant amplitude signal (say, 
+1) is transmitted in SSK. Therefore, the number of information bits 
conveyed in one channel use in SSK is $\lfloor\log_2n_t\rfloor$. 
For example, for $n_t=4$, the two-bit combinations $\{00,01,10,11\}$ are 
mapped to antenna indices $\{1,2,3,4\}$; antenna 1 is activated when input 
bits are 00, antenna 2 is activated when input bits are 01, antenna 3 
is activated when input bits are 10, and antenna 4 is activated when
input bits are 11. The problem of SSK signal detection at the receiver 
in a given channel use then becomes one of finding which one among the 
$n_t$ antennas was activated, i.e., determining the index of the active 
antenna. Assuming that the choice of an active antenna among all antennas 
is equally likely and a `+1' was sent on the active antenna, and that the 
channel gains from $n_t$ transmit antennas to $n_r$ receive antennas are 
known at the receiver, the maximum likelihood (ML) decision rule to find 
the active antenna index is given by
\begin{eqnarray}
\vspace{-2mm}
\hat{j} & = & {\arg\min_{j, \,\, 1\leq j \leq n_t}} \quad \| {\bf y} - {\bf h}_j\|^2,
\label{ssk_ml}
\end{eqnarray}
where ${\bf y}$ is the $n_r\times 1$ received signal vector and ${\bf h}_j$
is the $n_r\times 1$ channel gain vector from transmit antenna $j$ to the
receive antennas. The estimated antenna index $\hat{j}$ is then demapped 
to the information bits which represent that index. 
In SM signal detection, in addition to detecting the active antenna index,
information bits conveyed through the conventional modulation symbol carried
on the active antenna also have to be detected. 

A lot of recent research has focused on SM and SSK in point-to-point 
as well as cooperative relaying settings (see \cite{lajos},\cite{lh1} 
and the references therein). Bit error performance of SSK and SM 
in single-user point-to-point communication has been analyzed in 
\cite{renzo1},\cite{renzo2}. 
Transmit diversity schemes for SM MIMO (i.e., systems that combine SM 
and space-time coding) have been analyzed in \cite{renzo3}.
Multiuser SM MIMO on the downlink has been analyzed in \cite{lh2}.
SSK and SM employed on the uplink in 
multiuser MIMO systems have been studied in \cite{mu4}-\cite{mu5}. In 
\cite{mu2}-\cite{mu5}, it has been shown that, for the same spectral 
efficiency, multiuser SM-MIMO can outperform conventional multiuser MIMO. 
In this paper, we are interested in a modulation scheme which is a 
generalization of SM, referred to as {\em generalized spatial modulation
(GSM)} \cite{gsm0},\cite{gsm1}. GSM was introduced in 
\cite{gsm0} and its achievable rate was studied in detail in \cite{gsm1}.
Here, we are interested in the performance analysis and signal detection 
of multiuser GSM on the uplink in large-scale MIMO systems. Such a study 
has not been reported in the literature before. 

In GSM, the number of transmit RF chains, $n_{rf}$, is parameterized such 
that $1\leq n_{rf} \leq n_t$, and, in a given channel use, $n_{rf}$ out of 
$n_t$ transmit antennas are chosen and activated 
\cite{gsm0},\cite{gsm1},\cite{book}. The remaining $n_t-n_{rf}$ antennas 
remain silent. On the chosen antennas, $n_{rf}$ modulation symbols (one on 
each chosen antenna) are transmitted. The indices of the $n_{rf}$ active
antennas out of $n_t$ available antennas convey 
$\lfloor \log_2{n_{t}\choose n_{rf}} \rfloor$ information bits. This is
in addition to the $n_{rf}\lfloor \log_2|{\mathbb A}|\rfloor$ information 
bits conveyed by the $n_{rf}$ modulation symbols. It can be seen that 
both SM and spatial multiplexing turn out to be special cases of GSM 
with $n_{rf}=1$ and $n_{rf}=n_t$, respectively. In \cite{gsm1}, it has 
been shown that for a given modulation alphabet and $n_t$, there exists 
an optimum $n_{rf}$ in GSM that maximizes the spectral efficiency. 

In this paper, we consider the uplink in multiuser MIMO systems, where 
the BS has a large number of receive antennas (tens to hundreds) and each 
user terminal employs GSM with $n_t$ transmit antennas and $n_{rf}$ 
transmit RF chains. Some GSM configurations of interest at the user 
terminal include: ($n_t=4$, $n_{rf}=2$), ($n_t=8$, $n_{rf}=2$). When
$n_t>1$ and $n_{rf}=1$, GSM specializes to SM; example, ($n_t=2$, $n_{rf}=1$), 
($n_t=4$, $n_{rf}=1$). When $n_t=1$ and $n_{rf}=1$, GSM specializes
to conventional modulation. Our contributions in this paper can be 
summarized as follows.
\begin{itemize}
\item 	We first analyze the average bit error probability (ABEP) of
	multiuser GSM-MIMO under maximum-likelihood (ML) detection.
	We derive an upper bound on the ABEP, which is
	tight at moderate to high signal-to-noise ratios (SNR).
\item	We then propose low-complexity algorithms for GSM-MIMO signal 
	detection and channel estimation at the BS receiver based on 
	message passing. The proposed receiver algorithms scale very
	well in complexity and achieve near-ML performance in large
	dimensions. Simulation results show that, for the same spectral
	efficiency, multiuser GSM-MIMO can outperform multiuser SM-MIMO 
	as well as conventional multiuser MIMO, by about 2 to 9 dB at a 
	bit error rate (BER) of $10^{-3}$. Such SNR gains in GSM-MIMO 
	compared to SM-MIMO and conventional MIMO can be attributed to 
	the fact that, because of a larger number of spatial index bits, 
	GSM-MIMO can use a lower-order QAM alphabet which is more power 
	efficient.
\item	We carry out a study of the proposed and existing algorithms, 
	which includes an assessment of their performance and computational 
	complexity. Simulation results show that the proposed detection 
	algorithms have lesser complexity than minimum mean square error 
	(MMSE) detection complexity, while achieving significantly better 
	performance than MMSE detection performance.
\end{itemize}
The rest of the paper is organized as follows. The multiuser GSM-MIMO 
system model on the uplink is presented in Section \ref{sec2}. In 
Section \ref{sec3}, we derive an analytical upper bound on the ML 
detection performance in multiuser GSM-MIMO. In Section \ref{sec4}, 
we present the proposed detection and channel estimation algorithms
for multiuser GSM-MIMO and their performance. In Section \ref{sec5}, 
the performance of the proposed algorithms in frequency-selective 
fading are presented. Conclusions are presented in Section \ref{sec6}.

\begin{figure}[t]
\centering
\includegraphics[height=1.6in]{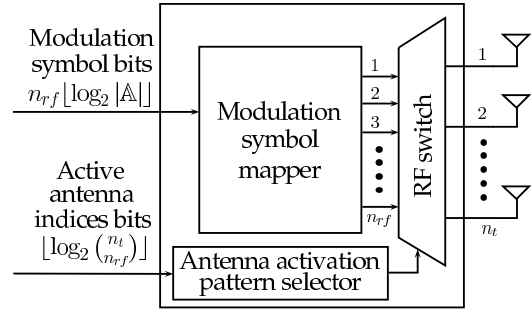}
\caption{GSM transmitter.}
\label{sys}
\vspace{-3mm}
\end{figure}

\section{Multiuser GSM-MIMO system model}
\label{sec2}
Consider a multiuser system with $K$ uplink users communicating with 
a BS having $N$ receive antennas, where $N$ is in the order of tens 
to hundreds. The ratio $K/N$ is the system loading factor. 
Users employ GSM for their transmission. Each user has $n_t$ transmit 
antennas and $n_{rf}$, $1\leq n_{rf} \leq n_t$, transmit RF chains. 
An $n_{rf}\times n_t$ switch connects the RF chains 
to the transmit antennas.
In a given channel use, each user selects $n_{rf}$ of its $n_t$ 
transmit antennas, and transmits $n_{rf}$ symbols from a modulation 
alphabet $\sa$ on the selected antennas. 
The remaining $n_t-n_{rf}$ antennas remain silent (i.e., they can be
viewed as transmitting the value zero). 
The GSM transmitter at the user terminal is shown in Fig. \ref{sys}. 
The selection of $n_{rf}$ active antennas is made based on 
$\lfloor\log_2{n_t \choose n_{rf}}\rfloor$ information bits. For example, 
for $n_t=4$ and $n_{rf}=2$, two out of the four antennas are selected using 
$\lfloor\log_2{4 \choose 2}\rfloor = \lfloor\log_2{6}\rfloor = 
\lfloor 2.585 \rfloor = 2$ information bits. The mapping of information 
bits to active antenna indices in GSM is described below. 

Define an `antenna activation pattern' to be an $n_t\times 1$ vector 
consisting of 1's and 0's, where a 1 in a coordinate indicates that 
the antenna corresponding to that coordinate is active and a 0 indicates 
that the corresponding antenna is silent. Note that there are 
$n_t\choose n_{rf}$ activation patterns possible. For example, for 
$n_t=4$ and $n_{rf}=2$, the following six activation patterns are 
possible:

\vspace{-4mm}
{\small
\[
[1 \,\, 1 \,\, 0\,\, 0]^T\hspace{-0.5mm}, 
[0\,\, 0\,\, 1\,\, 1]^T\hspace{-0.5mm},
[1\,\, 0\,\, 1\,\, 0]^T\hspace{-0.5mm}, 
[0\,\, 1\,\, 0\,\, 1]^T\hspace{-0.5mm},
[1\,\, 0\,\, 0\,\, 1]^T\hspace{-0.5mm}, 
[0\,\, 1\,\, 1\,\, 0]^T\hspace{-1mm}.
\]}
Out of the $n_t\choose n_{rf}$ possible activation patterns, only 
$2^{\lfloor \log_2{n_t\choose n_{rf}}\rfloor}$ activation patterns 
are needed for signaling. Let $\sap$ denote the set of these
$2^{\lfloor \log_2{n_t\choose n_{rf}}\rfloor}$ activation patterns
chosen from the set of all possible patterns,
i.e., $|\sap| = 2^{\lfloor \log_2{n_t\choose n_{rf}}\rfloor}$. 
In the above example, 
let the set of chosen activation patterns be
\[
\sap=\{[1 \,\, 1 \,\, 0\,\, 0]^T, [1\,\, 0\,\, 1\,\, 0]^T, [1\,\, 0\,\, 0\,\, 1]^T, [0\,\, 1\,\, 1\,\, 0]^T\}. 
\]
A mapping is done between combinations of
${\lfloor \log_2{n_t\choose n_{rf}}\rfloor}$ information bits to 
activation patterns in $\sap$. The following table shows such a 
mapping for the $n_t=4$, $n_{rf}=2$ example:
\vspace{-1mm}
\begin{table}[h]
\centering
\begin{tabular}{|c|c|c|}
\hline
Information bits & Antenna activation pattern & Remarks \\
\hline\hline
00 & $[1 \,\, 1 \,\, 0 \,\, 0]^T$ & antennas 1,2: Active;\\& & antennas 3,4: Silent\\ \hline
01 & $[1 \,\, 0 \,\, 1 \,\, 0]^T$ & antennas 1,3: Active;\\& & antennas 2,4: Silent \\ \hline
10 & $[1 \,\, 0 \,\, 0 \,\, 1]^T$ & antennas 1,4: Active;\\& & antennas 2,3: Silent \\ \hline
11 & $[0 \,\, 1 \,\, 1 \,\, 0]^T$ & antennas 2,3: Active;\\& & antennas 1,4: Silent \\ \hline
\end{tabular}
\caption{Mapping between information bits and active antenna indices
in GSM for $n_t=4$, $n_{rf}=2$.}
\vspace{-6mm}
\label{tab_map}
\end{table}

Note that the mapping does not need channel state information. Also, 
the mapping rule between information bits and active antenna indices 
is made known to the transmitter and receiver a priori for encoding 
and decoding purposes, respectively.

Apart from the bits conveyed through active antenna indices, additional
bits are conveyed through modulation symbols sent through the $n_{rf}$ 
active antennas. Therefore, the total number of bits conveyed by a 
GSM transmitter per channel use is given by
\vspace{-3mm}
\[
\Big\lfloor\log_2{n_t \choose n_{rf}}\Big\rfloor+n_{rf}\left\lfloor\log_2|\sa|\right\rfloor \quad \mbox{bpcu}.
\]
For example, a GSM transmitter with $n_t=4$, $n_{rf}=2$ and 4-QAM conveys
6 bpcu.

{\em GSM signal set:}
Let $\gsm$ denote the GSM signal set, which is the set of GSM 
signal vectors that can be transmitted. Then, $\gsm$ is given by
\begin{eqnarray}
\gsm = \big \{ \vs :  s_j \in \sa\cup\{0\}, \, \lVert\vs\rVert_0=n_{rf}, \, {\cal I}(\vs)\in\sap \big \},  
\end{eqnarray}
where $\vs$ is the $n_t\times 1$ transmit vector, $s_j$ is the $j$th 
entry of $\vs$, $j=1,\cdots,n_t$, $\lVert\vs\rVert_0$ is the $l_0$-norm of 
the vector $\vs$, and ${\cal I}(\vs)$ is a function that gives the activation 
pattern for $\vs$; for example, 
${\cal I}({\bf s}=[+1 \,\, -1 \,\, -1 \,\, 0 ]^T)=[1 \,\, 1 \,\, 1 \,\, 0]^T$. 

{\em Example:} Let $n_t=4$, $n_{rf}=2$, BPSK modulation, and 
$\sap=\{[1 \,\, 1 \,\, 0\,\, 0]^T, [1\,\, 0\,\, 1\,\, 0]^T, [1\,\, 0\,\, 0\,\, 1]^T, [0\,\, 1\,\, 1\,\, 0]^T\}$. 
The GSM signal set for these parameters is given by
{\scriptsize
\begin{eqnarray}
\hspace{-0mm}
{\mathbb S}_{4,\mbox{{\tiny BPSK}}}^2
\hspace{-3mm}&=&\hspace{-3mm}\left\{
\begin{bmatrix} +1 \\ +1 \\ 0 \\ 0\end{bmatrix}\hspace{-0.7mm},
\begin{bmatrix} +1 \\ -1 \\ 0 \\ 0\end{bmatrix}\hspace{-0.7mm},
\begin{bmatrix} -1 \\ -1 \\ 0 \\ 0\end{bmatrix}\hspace{-0.7mm},
\begin{bmatrix} -1 \\ +1 \\ 0 \\ 0\end{bmatrix}\hspace{-0.7mm},
\begin{bmatrix} +1 \\ 0 \\ +1 \\ 0\end{bmatrix}\hspace{-0.7mm},
\begin{bmatrix} +1 \\ 0 \\ -1 \\ 0\end{bmatrix}\hspace{-0.7mm},
\begin{bmatrix} -1 \\ 0 \\ -1 \\ 0\end{bmatrix}\hspace{-0.7mm},
\begin{bmatrix} -1 \\ 0 \\ +1 \\ 0\end{bmatrix}\hspace{-0.7mm}, \right.
\nonumber \\ & & \left.
\hspace{-0.5mm}
\begin{bmatrix} +1 \\ 0  \\ 0\\ +1\end{bmatrix}\hspace{-0.7mm},
\begin{bmatrix} +1 \\ 0  \\ 0\\ -1\end{bmatrix}\hspace{-0.7mm},
\begin{bmatrix} -1 \\ 0  \\ 0\\ -1\end{bmatrix}\hspace{-0.7mm},
\begin{bmatrix} -1 \\ 0  \\ 0\\ +1\end{bmatrix}\hspace{-0.7mm},
\begin{bmatrix} 0\\ +1 \\ +1 \\ 0\end{bmatrix}\hspace{-0.7mm},
\begin{bmatrix} 0\\ +1 \\ -1 \\ 0\end{bmatrix}\hspace{-0.7mm},
\begin{bmatrix} 0\\ -1 \\ -1 \\ 0\end{bmatrix}\hspace{-0.7mm},
\begin{bmatrix} 0\\ -1 \\ +1 \\ 0\end{bmatrix}
\right \}. \nonumber
\end{eqnarray}
}

\begin{figure*}
\subfigure[Multiuser GSM-MIMO system.]{
\includegraphics[width=3.15in,height=3.00in]{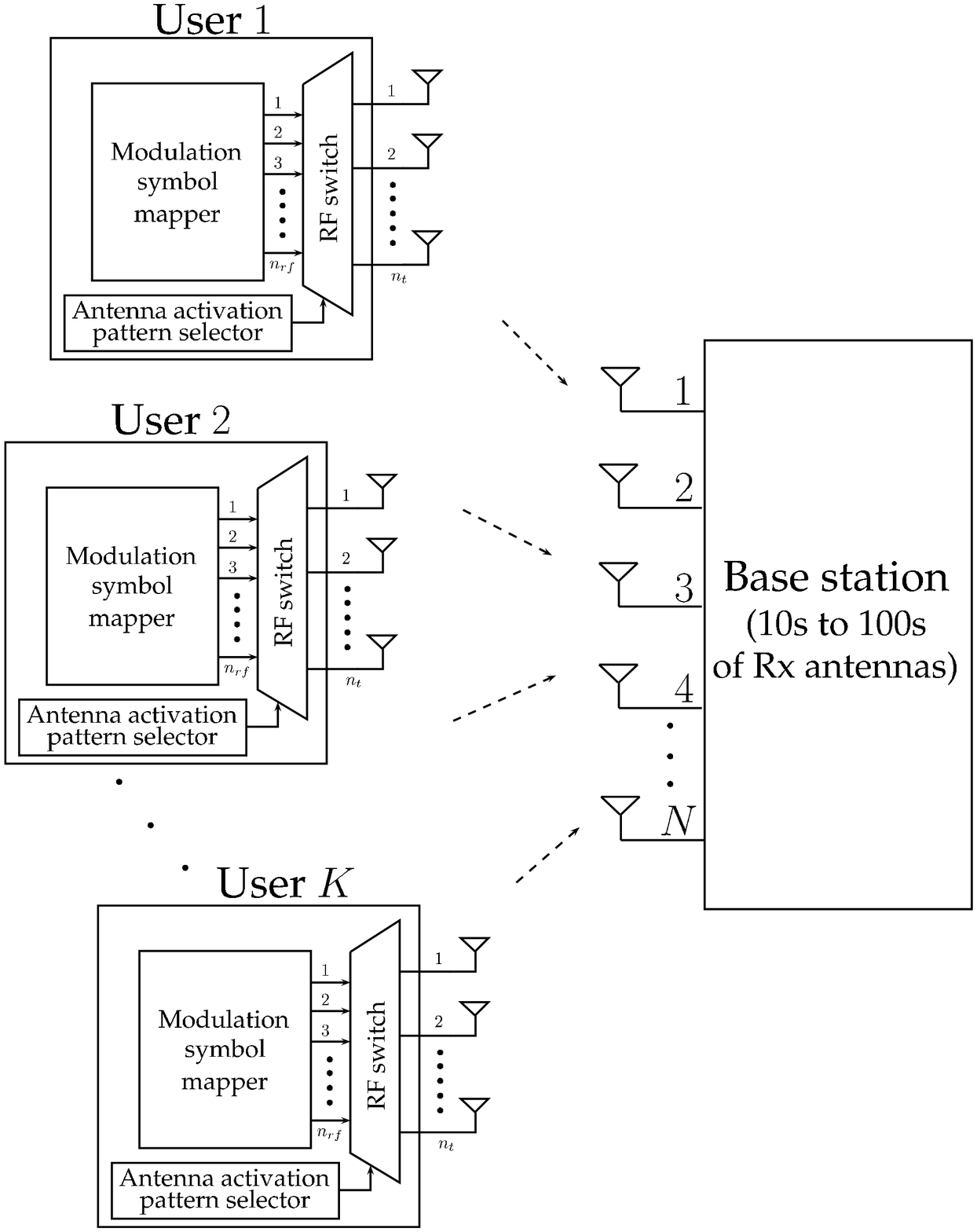}
\label{sys1}
}
\hspace{6mm}
\subfigure[Conventional multiuser MIMO system.]{ 
\includegraphics[width=3.15in,height=3.00in]{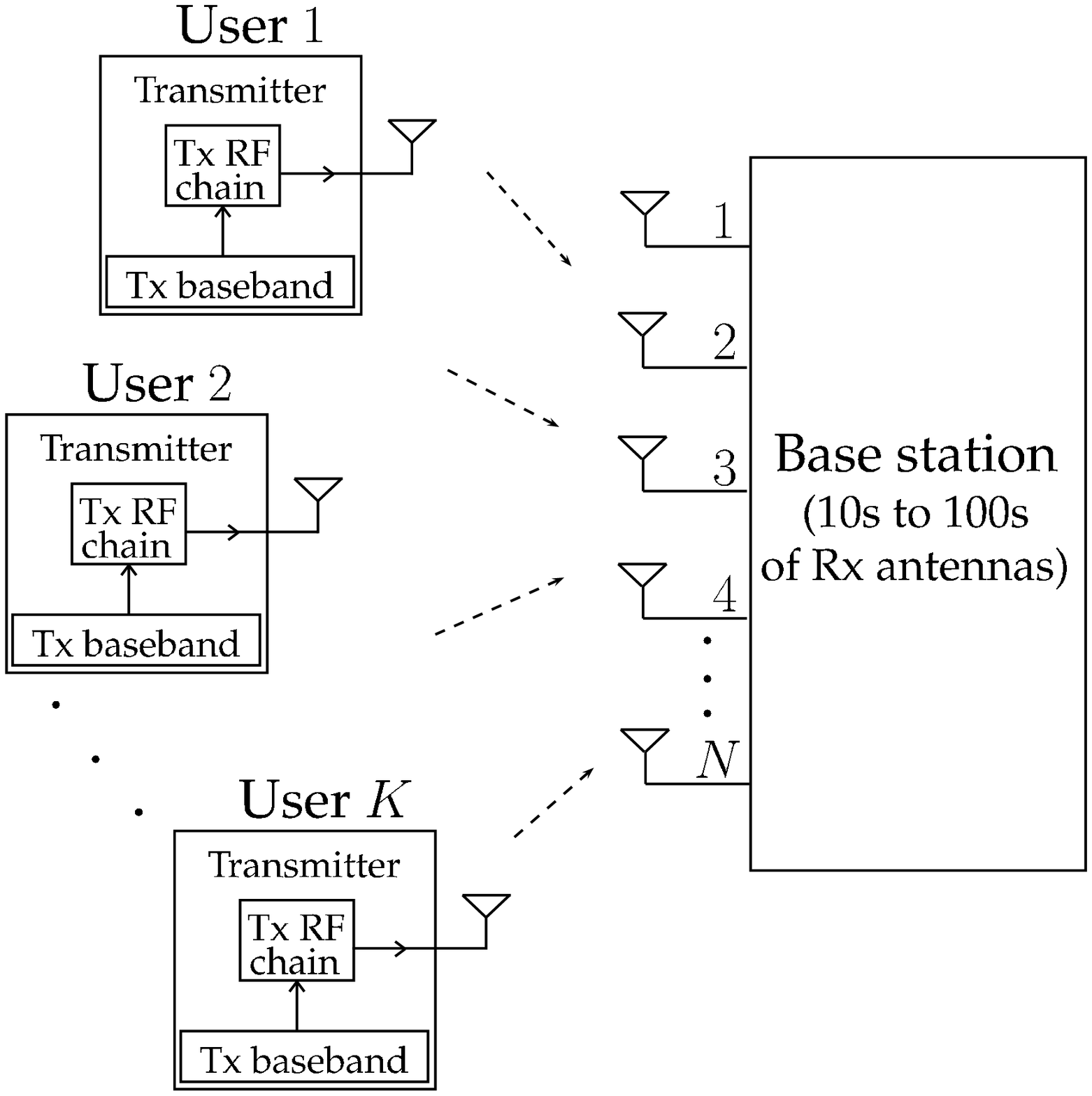}
\label{sys2}
}
\vspace{-2mm}
\caption{Large-scale multiuser GSM-MIMO and conventional multiuser
MIMO system architectures.}
\vspace{-3mm}
\end{figure*}

Figures \ref{sys1} and \ref{sys2} illustrate large-scale multiuser 
GSM-MIMO system and conventional multiuser MIMO (massive MIMO) system, 
respectively. Let $\vx_k \in \gsm$ denote the transmit vector 
from user $k$. Let
$\vx \Define [\vx_1^T \ \ \vx_2^T\,\cdots\,\vx_k^T\,\cdots\,\vx_K^T]^T$ 
denote the vector comprising of transmit vectors from all the users,
where $(.)^T$ denotes transpose operation. Note that $\vx \in (\gsm)^K$.

Let $\mh \in \mathbb{C}^{N\times Kn_t}$ denote the channel gain matrix,
where $H_{i,(k-1)n_t+j}$ denotes the complex channel gain from the $j$th 
transmit antenna of the $k$th user to the $i$th BS receive antenna. The 
channel gains are assumed to be independent Gaussian with zero mean and 
variance $\sigma_{\kappa}^2$, such that 
$\sum_{\kappa=1}^{Kn_t} \sigma_{\kappa}^2=Kn_t$. The $\sigma_{\kappa}^2$ 
models the imbalance in the received power from the ${\kappa}$th antenna, 
${\kappa} \in\{1,\cdots,Kn_t\}$, due to path loss etc., and 
$\sigma_{\kappa}^2=1$ corresponds to the case of perfect power control. 
Assuming perfect synchronization, the received signal at the $i$th BS
antenna is given by
\vspace{-2mm}
\begin{eqnarray}
y_i & \hspace{1mm}= &\hspace{1mm} \sum_{k=1}^{K} \vh_{i,[k]}\vx_k \hspace{1mm}+ \hspace{1mm}n_i,
\end{eqnarray}
where $\vh_{i,[k]}$ is a $1\times n_t$ vector obtained from the $i$th row of 
$\mh$ and $(k-1)n_t+1$ to $kn_t$ columns of $\mh$, and $n_i$ is the noise 
modeled as a complex Gaussian random variable with zero mean and variance 
$\sigma^2$. The received signal at the BS antennas can be written in vector 
form as
\begin{eqnarray}
\vy& = & \mh\vx+\vn,
\label{sysmodel}
\end{eqnarray}
where $\vy = [y_1, \ y_2, \cdots, \ y_N]^T$ and 
$\vn = [n_1, \ n_2, \cdots, \ n_N]^T$. 
For this system model, the ML detection rule is given by
\begin{equation} 
\label{ml} 
\hat{\vx}=\argmin_{\vx\in (\gsm)^K} \ \|\vy-\mh\vx\|^2,  
\end{equation}
where $\|\vy-\mh\vx\|^2$ is the ML cost.
The maximum a posteriori  probability (MAP) decision rule is given by
\begin{eqnarray}
\label{map}
\hat{\vx} = \argmax_{\vx\in (\gsm)^K} \ \Pr(\vx\mid\vy,\mh). 
\end{eqnarray}
Since $|(\gsm)^K|=(|\sap||\sa|^{n_{rf}})^K$,
the exact computation of (\ref{ml}) and (\ref{map}) requires exponential
complexity in $K$. In the next section, we derive an analytical upper
bound on the ABEP for ML detection. In Section \ref{sec4}, we propose 
message passing based detection algorithms which give approximate 
solutions to (\ref{map}) at low complexities. 

We note that the condition for the spectral efficiencies of GSM-MIMO 
(with $n_t$ transmit antennas, $n_{rf}$ transmit RF chains, and 
modulation alphabet ${\mathbb A}$ at each user) and conventional 
multiuser MIMO (with $m_t$ transmit antennas, $m_t$ transmit RF 
chains, and modulation alphabet ${\mathbb B}$ at each user) to be 
the same is given by
\[
|\sap||\sa|^{n_{rf}} = |{\mathbb B}|^{m_t}.
\]
For example, ($i$) GSM-MIMO with BPSK, $n_t=4$, $n_{rf}=2$, 
($ii$) SM-MIMO with 4-QAM, $n_t=4$, $n_{rf}=1$, and ($iii$)
conventional multiuser MIMO with 4-QAM, 2 transmit antennas and
2 transmit RF chains, all have the same spectral efficiency of 
4 bpcu per user.  In conventional multiuser MIMO, the vector 
$\vx \in {\mathbb B}^{Km_t}$ and the channel matrix 
$\mh \in {\mathbb C}^{N\times Km_t}$. 

\section{Average BEP Analysis}
\label{sec3}
In this section, we analyze the average BEP performance of ML detection 
in multiuser GSM-MIMO. Assume that all the transmit GSM signal vectors 
are equally likely. The ML detection rule in (\ref{ml}) can be written as 
\vspace{-1mm}
\begin{eqnarray} 
\hat{\vx} = \argmin_{\vx\in \gsmk} \ \|\vy-\sum_{\kappa=1}^{Kn_{t}} x_{\kappa}\vh_{\kappa}\|^2,
\vspace{-1mm}
\end{eqnarray}
where $x_{\kappa}$ is the $\kappa$th element of $\vx$, $\vh_{\kappa}$ is 
the $\kappa$th 
column of $\mh$ and $\gsmk\Define(\gsm)^K$. The pairwise error 
probability (PEP) of $\vx$ being decoded as $\tilde{\vx} \in \gsmk$ 
can be written as

\vspace{-2mm}
{\small
\begin{eqnarray} 
\vspace{-8mm}
P(\vx \rightarrow \tilde{\vx}|\mh)& \hspace{-2mm} = & \hspace{-2mm}
P\Big(\|\vy-\sum_{\kappa=1}^{Kn_{t}} x_{\kappa}\vh_{\kappa}\|^2 > \|\vy-\sum_{\kappa=1}^{Kn_{t}} \tilde{x}_{\kappa}{\vh}_{\kappa}\|^2\big|\mh\Big) \nonumber\\  
&\hspace{-35mm}=&\hspace{-19mm}
P\Big(\sum_{i=1}^{N}|y_i-\sum_{\kappa=1}^{Kn_{t}} x_{\kappa}h_{i,\kappa}|^2 > \sum_{i=1}^{N}|y_i-\sum_{\kappa=1}^{Kn_{t}} \tilde{x}_{\kappa}h_{i,\kappa}|^2\big|\mh\Big), 
\end{eqnarray}} 

\hspace{-5mm}
where $h_{i,\kappa}$ is the $i$th element of ${\bf h}_{\kappa}$.
Let $A_i \Define \sum_{\kappa=1}^{Kn_{t}} x_{\kappa}h_{i,\kappa}$ and 
$\tilde{A_i} \Define \sum_{\kappa=1}^{Kn_{t}} \tilde{x}_{\kappa}h_{i,\kappa}$.
Since
$\vx$ is the transmitted vector, $y_i = A_i + n_i$, $i=1,\cdots,N$.  
Now, we can write
{\small
\begin{eqnarray}
P(\vx\rightarrow\tilde{\vx}|\mh)&\hspace{-2mm}=&\hspace{-2mm}P\Big(\sum_{i=1}^{N}|y_i-A_i |^2 > \sum_{i=1}^{N}|y_i-\tilde{A_i}|^2\big|\mh\Big)
 \nonumber \\
&\hspace{-22mm}=&\hspace{-12mm}P\Big(\sum_{i=1}^{N}|n_i |^2 > \sum_{i=1}^{N}|A_i+n_i-\tilde{A_i}|^2\big|\mh\Big)
 \nonumber \\ 
&\hspace{-22mm}=&\hspace{-12mm}P\Big(\sum_{i=1}^{N}2\Re((\tilde{A_i}-A_i)n_i^{*}) > \sum_{i=1}^{N}|A_i-\tilde{A_i}|^2\big|\mh\Big), 
\end{eqnarray}}
\vspace{-1mm}

\hspace{-5mm}
where $\Re(.)$ denotes real part, $(.)^*$ denotes conjugation, and 
$\sum_{i=1}^{N}2\Re((\tilde{A_i}-A_i)n_i^{*})$ is a Gaussian random 
variable with mean zero and variance 
$2\sigma^2\sum_{i=1}^{N}|A_i-\tilde{A_i}|^2$. Therefore, 

\vspace{-2mm}
{\small
\begin{eqnarray}
\hspace{-4mm}
P(\vx\rightarrow\tilde{\vx}|\mh)&\hspace{-2mm} = & \hspace{-2mm} Q\Bigg(\sqrt{\sum_{i=1}^{N}|A_i-\tilde{A_i}|^2/2\sigma^2}\, \,\Bigg)
\nonumber \\
&\hspace{-2mm}=& \hspace{-2mm} Q\Bigg(\sqrt{\Big\|\sum_{\kappa=1}^{Kn_{t}}(x_{\kappa}-\tilde{x}_{\kappa}){\vh}_{\kappa}\Big\|^2/2\sigma^2}\, \,\Bigg).
\label{qf}
\end{eqnarray}}
The argument in (\ref{qf}) has a central $\chi^2$-distribution with $2N$ 
degrees of freedom. Computation of the unconditional PEP 
$P(\vx\rightarrow\tilde{\vx})$ requires the expectation of the $Q(.)$ 
function in (\ref{qf}) w.r.t. ${\bf H}$, which can be obtained as follows 
\cite{anl}:
\begin{eqnarray}
\hspace{-4mm}
{\text{$P$}}(\vx\rightarrow\tilde{\vx})&\hspace{-2mm}=&\hspace{-2mm}\mathbb{E}_{\mh}\{ P(\vx\rightarrow\tilde{\vx}|\mh)\}\nonumber \\
&\hspace{-2mm}=&\hspace{-2mm} f(\alpha)^{N}\sum_{i=0}^{N-1}{N-1+i\choose i}(1-f(\alpha))^i,
\label{pepexpr}
\end{eqnarray}
where 
$f(\alpha) \Define \frac{1}{2}\Big(1-\sqrt{\frac{\alpha}{1+\alpha}}\,\Big)$,
$\alpha \Define\frac{1}{4\sigma^2} \sum\limits_{\kappa=1}^{Kn_{t}}\theta_{\kappa}$, 
and $\theta_{\kappa} \Define |x_{\kappa}-\tilde{x}_{\kappa}|^2$.
Now, an upper bound on the average BEP based on union bounding can be 
obtained as
\begin{eqnarray}
P_{B} &\hspace{-2mm} \leq & \hspace{-2mm}  
\frac{1}{2^{\eta}} \sum_{\vx\in \gsmk} \,\, \sum_{\tilde{\vx}\in \gsmk\setminus\vx} P(\vx\rightarrow\tilde{\vx})\frac{d(\vx,\tilde{\vx})}{\eta}, \label{qq1}
\end{eqnarray}
where $d(\vx,\tilde{\vx})$ is the number of bits in which $\vx$ differs 
from $\tilde{\vx}$. The total number of PEPs to be calculated in 
(\ref{qq1}) is $2^\eta(2^\eta-1)$. Therefore, the complexity of the 
computation of the bound in (\ref{qq1}) will increase exponentially 
in $K, n_t$, and $n_{rf}$. In the following subsection, we devise 
simplification methods to reduce this computational complexity. 

\vspace{-2mm}
\subsection{Reduction of computation complexity in (\ref{qq1})}
The expression for $P_B$ in (\ref{qq1}) can be written in the following 
form: 
\vspace{-1mm}
{\small
\begin{eqnarray}
\hspace{-1mm}
P_{B} &\hspace{-2mm} \leq & \hspace{-2mm} \frac{1}{2^\eta}\sum_{i=1}^{|{\cal S}^K|}\sum_{j=1}^{|{\cal S}^K|}\sum_{\vx:{\cal I}(\vx)={\bf s}_i\in\sap^K}\,\,\sum_{\substack{\tilde{\vx}:{\cal I}(\tilde{\vx})={\bf s}_j\in\sap^K, \\ \tilde{\vx}\neq\vx}}\hspace{-6mm}P(\vx\rightarrow\tilde{\vx})\frac{d(\vx,\tilde{\vx})}{\eta}.\nonumber \\
\label{bnd}
\end{eqnarray}
}
\vspace{-6mm}

\hspace{-5mm}
For a given pair of antenna activation patterns ${\bf s}_i$ and ${\bf s}_j$, 
$i,j\in\{1,\cdots,|\sap|^K\}$, the total number of PEPs are 
$|\sa|^{2Kn_{rf}}$ when $i\neq j$, and 
$|\sa|^{Kn_{rf}}(|\sa|^{Kn_{rf}}-1)$ when $i=j$. 

{\em Complexity reduction 1}: For a pair of activation patterns ${\bf s}_i$ 
and ${\bf s}_j$, let ${\mathcal A}_{ij}$ denote the set of active antennas 
that are common to both ${\bf s}_i$ and ${\bf s}_j$. Define 
$\beta_{ij}= Kn_{rf} - |{\mathcal A}_{ij}|$. Note that 
$\beta_{ij} \in\{0,1,\cdots,\min(n_{rf},n_t-n_{rf})K\}$. 
Also, note that for any $i,j$ for which $\beta_{ij}=q$, 
the value of the summation
$\sum\limits_{\vx:{\cal I}(\vx)={\bf s}_i}\,\sum\limits_{\substack{\tilde{\vx}:{\cal I}(\tilde{\vx})={\bf s}_j, \, \tilde{\vx}\neq\vx}}\hspace{-3mm}P(\vx\rightarrow\tilde{\vx})$
in (\ref{bnd}) will be the same, and so it is enough to compute this
summation only once for each $q$. With this simplification, (\ref{bnd}) 
can be written as

\vspace{-3mm}
{\small 
\begin{eqnarray*}
\hspace{-2mm}
P_{B} & \hspace{-2mm} \leq & \hspace{-2mm} \frac{1}{2^{\eta}}\sum_{q=0}^{\min(n_{rf},n_t-n_{rf})K}\hspace{-7mm}\phi(q)\sum_{\vx:{\cal I}(\vx)={\bf s}_i}\,
\sum_{\substack{\tilde{\vx}:{\cal I}(\tilde{\vx})={\bf s}_j\\\beta_{ij}=q}}\hspace{-3mm}P(\vx\rightarrow\tilde{\vx})\frac{d(\vx,\tilde{\vx})}{\eta},
\label{qq2}
\end{eqnarray*}}
\vspace{-3mm}

\hspace{-5mm}
where $\phi(q)$ is the number of $({\bf s}_i,{\bf s}_j)$ pairs for
which $\beta_{ij}=q$, which can be computed easily. 

{\em Example:}
A direct computation of the first two summations in (\ref{bnd}) is
prohibitive, as the total number of terms is exponential in $K, n_t$,
$n_{rf}$. For $K=2, n_t=4$ and $n_{rf}=2$, $|{\cal S}^K|=2^{4}$.
So, the first two summations in (\ref{bnd}) will have $2^{8}=256$ terms.
Whereas for these parameters, $q\in \{0, 1, 2, 3, 4\}$,
$\phi(\{0, 1, 2, 3, 4\})=\{16, 88, 128, 22, 2\}$ and
$\sum_{q=0}^{4}\phi(q)=256$. Hence, the inner summations can be
computed only 5 times (once for each $q$), instead of 256 times.

{\em Complexity reduction 2}:
For each value of $q$, we need to compute $|\sa|^{2Kn_{rf}}$ PEPs. 
We propose to reduce this complexity as follows. The parameter $\alpha$ 
in (\ref{pepexpr}) is the summation of $Kn_t$ terms.  Out of these $Kn_t$ 
terms, $Kn_t-(Kn_{rf}+q)$ terms will be zero for a given 
value of $q$. Of the $(Kn_{rf}+q)$ non-zero terms, $2q$ terms will take 
values from $\sj\Define\{|c|^2 : c\in\sa\}$, and $Kn_{rf}-q$ terms will 
take values from $\sL\Define\{|c-\tilde c|^2 : c, \tilde c\in\sa \}$. 
Let $\sj=\{j_1,j_2,\cdots,j_m\}$ and $\sL=\{l_1,l_2,\cdots,l_n\}$, where
$j_1 < j_2 < \cdots < j_m$,  $l_1 < l_2< \cdots < l_n$, $m=|\sj|$, and 
$n=|\sL|$. We write $\alpha$ as $\alpha=\alpha_1+\alpha_2$, where 
$\alpha_1$ is the sum of $2q$ terms from $\sj$ and $\alpha_2$ is the 
sum of $Kn_{rf}-q$ terms from $\sL$. 
Note that $\alpha_1$ can take values in the range 
$2qj_1$ to $2qj_m$. 
For a given value of $\alpha_1$, the following equations must be 
satisfied:
\begin{equation}
\sum_{i=1}^{m} j_iv_i = \alpha_1, \quad \quad
\sum_{i=1}^{m} v_i =2q,
\end{equation}
where $v_i$ is an integer such that 
$v_i\in \{0,1,\cdots,\lfloor(\alpha_1-\sum_{k=i+1}^{m}j_kv_k)/j_i\rfloor\}$. 
Similarly, $\alpha_2$ can take values in the range
$(Kn_{rf}-q)l_1$ to $(Kn_{rf}-q)l_n$, and, for a given value of $\alpha_2$, 
the following equations must be satisfied: 
\vspace{-1mm}
\begin{equation}
\sum_{i=1}^{n} l_iu_i = \alpha_2, \quad \quad
\sum_{i=1}^{n} u_i =Kn_{rf}-q,
\end{equation}
where $u_i$ is an integer such that
$u_i\in \{0,1,\cdots,\lfloor(\alpha_2-\sum_{k=i+1}^{n}l_ku_k)/l_i\rfloor\}$.
Since $\alpha=\alpha_1+\alpha_2$, $\alpha$ lies in the range 
$2qj_1+(Kn_{rf}-q)l_1$ to $2qj_m+(Kn_{rf}-q)l_n$. A given value of 
$\alpha$ can be written as
\begin{equation}
\alpha=\sum_{i=1}^{m} j_iv_i +\sum_{i=1}^{n} l_iu_i \quad \text{s.t.} \,
\sum_{i=1}^{m} v_i =2q, \sum_{i=1}^{n} u_i =Kn_{rf}-q.
\end{equation}
The choices of $v_i$'s and $u_i$'s to attain a particular $\alpha$ is not
unique, i.e., there exist multiple pairs of $\vx$ and ${\tilde \vx}$
that correspond to different values of $v_i$'s and $u_i$'s but the same
value of $\alpha$. Thus, we need to evaluate (\ref{pepexpr}) only once for 
a given  $\alpha$ and count the number of possible combinations of $v_i$'s 
and $u_i$'s that correspond to that $\alpha$. 

{\em Example:} When $n_t$ = 4, $n_{rf}$ = 3, 
$\sa = \{-1-{\bf j},-1+{\bf j},1-{\bf j},1+{\bf j}\}$, where 
${\bf j}=\sqrt{-1}$, then, $\sj = \{2\}$, $\sL = \{0,4,8\}$. For a 
particular value of $q$, say $q=1$, the summation in (\ref{qq2}) 
requires computation of PEP for 64 different pairs of GSM signals. 
But $\alpha$ lies in the range 2 to 18, and hence we need to compute 
only 17 PEP terms.   

\begin{figure}
\hspace{-4mm}
\includegraphics[width=3.75in,height=2.75in]{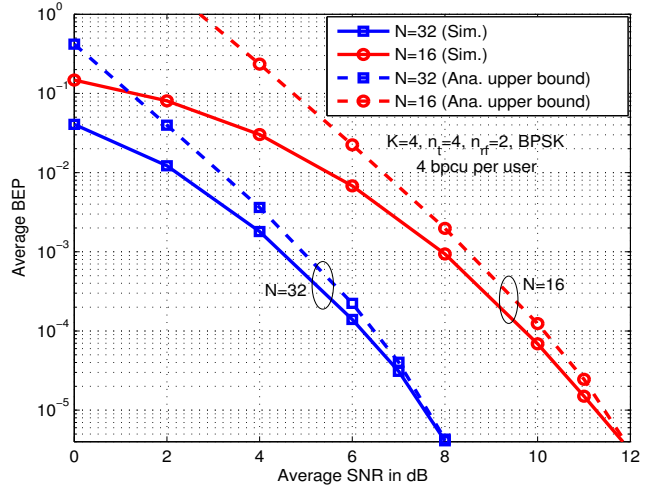}
\vspace{-8mm}
\caption{Comparison between analytical ABEP upper bound and simulated
ABEP for ML detection in GSM-MIMO with $N=16,32$, $K=4$, $n_t=4$, 
$n_{rf}=2$, BPSK, and 4 bpcu per user. Analysis and simulation.}
\vspace{-4mm}
\label{fig1}
\end{figure}

\vspace{-0mm}
\subsection{Numerical results}
\label{subsec_new}
In Fig. \ref{fig1}, we compare the analytical ABEP upper bound and the
simulated ABEP of multiuser GSM-MIMO with ML detection for the following 
system parameter settings: $N=16,32$, $K=4$, $n_t=4$, $n_{rf}=2$, BPSK, 
and 4 bpcu per user. It can be observed that the upper bound is tight 
at moderate to high SNRs. It is also observed that, as expected, both 
analysis and simulation predict that the ABEP performance improves as 
the number of BS antennas $N$ is increased.

\begin{figure}
\hspace{-4mm}
\includegraphics[width=3.75in,height=2.75in]{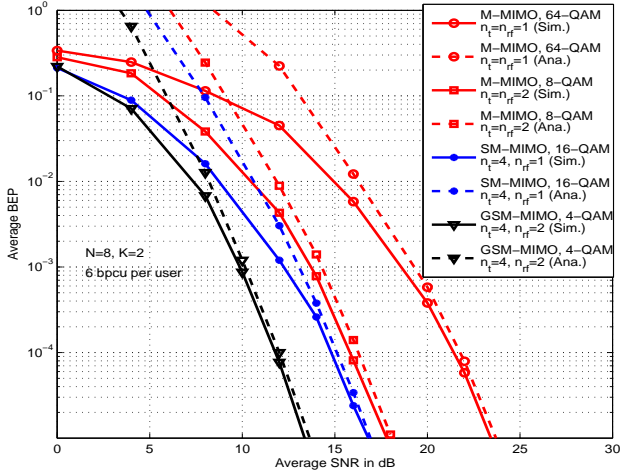}
\vspace{-10mm}
\caption{Comparison between the ABEP performance of four different
systems, all with $N=8$, $K=2$, and 6 bpcu per user: 
($i$) conventional multiuser MIMO, $n_t=n_{rf}=1$, 64-QAM;
($ii$) conventional multiuser MIMO, $n_t=n_{rf}=2$, 8-QAM;
($iii$) multiuser SM-MIMO, $n_t=4$, $n_{rf}=1$, 16-QAM; 
and ($iv$) multiuser GSM-MIMO, $n_t=4$, $n_{rf}=2$, 4-QAM. 
Analysis and simulation.}
\vspace{-4mm}
\label{fig1a}
\end{figure}

In Fig. \ref{fig1a}, we compare the ABEP performance of the following
four different systems with $N=8$ and $K=2$: 
System 1 -- conventional multiuser MIMO with $n_t=n_{rf}=1$, 16-QAM;
System 2 -- conventional multiuser MIMO with $n_t=n_{rf}=2$, 8-QAM;
System 3 -- multiuser SM-MIMO with $n_t=4$, $n_{rf}=1$, 16-QAM; 
and System 4 -- multiuser GSM-MIMO with $n_t=4$, $n_{rf}=2$, 4-QAM. 
Note that all the four systems achieve the same spectral efficiency 
of 6 bpcu per user. The first two systems are conventional multiuser
MIMO systems where $n_t=n_{rf}$. System 1 uses one transmit antenna
and one transmit RF chain at each user and achieves 6 bpcu per user 
by using 64-QAM. On the other hand, System 2 uses two transmit antennas 
and two transmit RF chains at each user and achieves 6 bpcu per user by 
using 8-QAM. System 3 is a multiuser SM-MIMO system where each user uses 
four transmit antennas but only one transmit RF chain. Each user in this 
system uses 16-QAM to achieve 6 bpcu per user; 4 bits through 16-QAM and 
2 bits through indexing. System 4 is a GSM-MIMO system where each user 
uses four transmit antennas and two transmit RF chains. This system uses 
4-QAM on two streams to achieve 6 bpcu per user; four bits through 
modulation symbols (i.e., two 4-QAM symbols on two streams)  and 2 bits 
through indexing. 

The following observations can be made from Fig. \ref{fig1a}. First, it 
can be observed that multiuser SM-MIMO system (System 3) achieves better 
performance compared to conventional multiuser MIMO systems (Systems 1 
and 2) by about 2 to 7 dB at $10^{-3}$ BER. The better performance of SM-MIMO 
over conventional MIMO in System 1 can be attributed to the fact that, to 
achieve the same spectral efficiency of 6 bpcu per user, SM-MIMO can use 
a smaller-sized QAM alphabet (16-QAM) than that used in system 1 (64-QAM), 
and that a smaller-sized QAM alphabet is more power efficient than a 
larger-sized one. Also, the better performance of SM-MIMO over
conventional MIMO in system 2 is attributed to the fact that, although
System 2 uses a smaller-sized alphabet (8-QAM) than that in SM-MIMO
(16-QAM), there is interference from multiple spatial streams in System 2.
Another observation in Fig. \ref{fig1a} is that multiuser GSM-MIMO system 
(System 4) 
performs better than multiuser SM-MIMO system (System 3) as well as 
conventional MIMO system in System 2 which also uses two RF chains
like GSM-MIMO. This is because, though GSM-MIMO uses two spatial streams
like conventional MIMO, its alphabet size is smaller than that in 
conventional MIMO.

\section{Low-complexity Receiver Algorithms} 
\label{sec4}
Optimal detection of multiuser GSM-MIMO signals in large dimensions is 
prohibitively complex. Generalized sphere decoding (GSD) approach 
\cite{gsd1},\cite{gsd2} can be employed for GSM-MIMO detection. But the 
complexity of GSD can be high (exponential complexity at low-to-medium 
SNRs). Therefore, for large systems, GSD becomes prohibitively complex.
In this section, we present low-complexity
near-optimal detection and channel estimation algorithms for large-scale 
multiuser GSM-MIMO systems. Two approximate message passing based 
algorithms for detection are presented. In the first algorithm, 
abbreviated as MP-GSM (message passing for GSM) detection algorithm, 
messages are formed based on the basic multiuser GSM-MIMO system model 
in (\ref{sysmodel}). In the second algorithm, abbreviated as CHEMP-GSM 
(channel hardening-exploiting message passing for GSM) algorithm, 
messages are formed based on a matched filtered version of the basic
system model in (\ref{sysmodel}). We also present a channel estimation 
approach that directly obtains an estimate of ${\bf H}^H{\bf H}$ for use 
in the CHEMP-GSM algorithm. 

\vspace{-2mm}
\subsection{MP-GSM detection algorithm}
\label{subsec4a}
\begin{figure}
\centering 
\subfigure[Factor graph]{
\includegraphics[width=2.75in,height=1.25in]{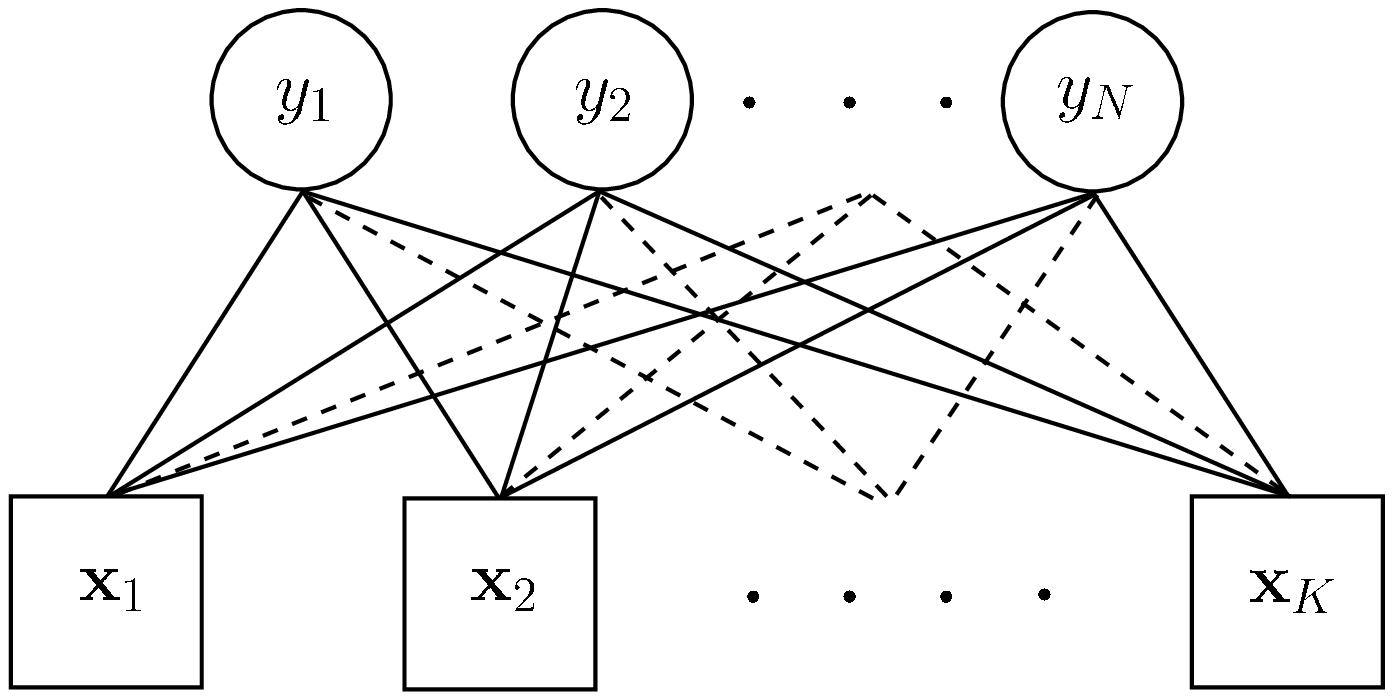}
\label{graph}
}
\subfigure[Observation node messages]{
\includegraphics[width=1.3in,height=1in]{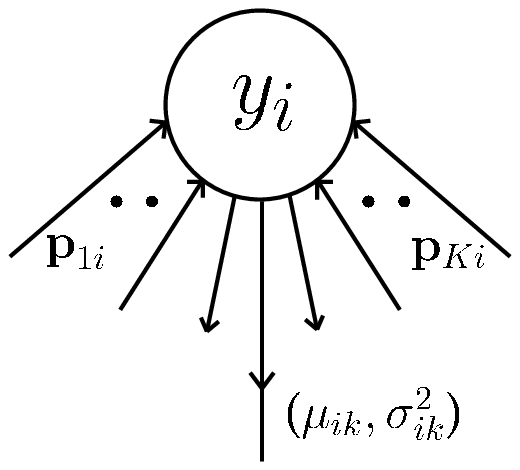}
\label{obs}
}
\subfigure[Variable node messages]{
\hspace{4mm}
\includegraphics[width=1.5in,height=0.9in]{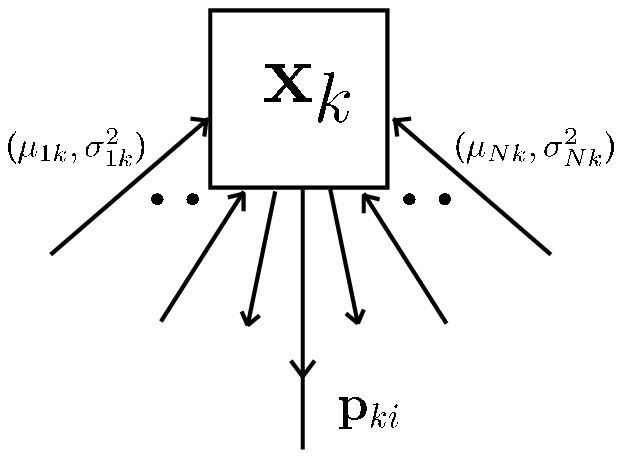}
\label{var}
}
\caption{The factor graph and messages passed in the MP-GSM algorithm.}
\vspace{-4mm}
\end{figure}

Consider the multiuser GSM-MIMO system model in (\ref{sysmodel}). We model 
this system as a fully connected factor graph with $K$ variable nodes (or 
factor nodes) corresponding to $\vx_k$'s and $N$ observation nodes 
corresponding to $y_i$'s, as shown in Fig. \ref{graph}. We aim to get 
approximate MAP solution through message passing on this graph, where 
messages are formed by approximating the probability density of the 
interference as Gaussian.

{\em Messages:}
The messages passed between variable nodes and observation nodes in
the factor graph are derived as follows. Equation (\ref{sysmodel}) 
can be written as
\vspace{-2mm}
\begin{eqnarray}
y_i & \hspace{-2mm} = & \hspace{-2mm} {\bf h}_{i,[k]}\vx_k+
\underbrace{\sum_{j=1, j\neq k}^K{\bf h}_{i,[j]}\vx_j + n_i}_{\Define \ g_{ik}},
\label{usreqn}
\end{eqnarray}
where ${\bf h}_{i,[j]}$ is a row vector of length $n_t$, given by 
$[H_{i,(j-1)n_t+1} \quad H_{i,(j-1)n_t+2} \, \cdots \, H_{i,jn_t}]$, 
and $\vx_j \in \gsm$.
The term $g_{ik}$ defined in (\ref{usreqn}) is approximated to be a 
Gaussian random variable\footnote{This Gaussian approximation will be 
accurate for large $K$; example, in systems with tens of users.} with mean 
$\mu_{ik}$ and variance $\sigma^2_{ik}$. The mean $\mu_{ik}$ in the
approximation is given by
\begin{eqnarray}
\mu_{ik}&\hspace{-2.0mm} = & \hspace{-2.0mm} \E\bigg[\sum_{j=1, j\neq k}^K \hspace{-2mm} {\bf h}_{i,[j]}\vx_j + n_i\bigg] \nonumber \\
& \hspace{-2mm} = &\hspace{-2mm} \sum_{j=1, j\neq k}^K \ \sum_{\vs\in\gsm} \hspace{-2mm} p_{ji}(\vs){\bf h}_{i,[j]}\vs \nonumber\\
&\hspace{-2mm} = & \hspace{-2mm} \sum_{j=1, j\neq k}^K \ \sum_{\vs\in\gsm}\hspace{-1mm}p_{ji}(\vs) \sum_{l\in{\cal I}(\vs)}\hspace{-1mm}s_{l} H_{i,(j-1)n_t+l},
\label{mu}
\end{eqnarray} 
where $s_{l}$'s are the non-zero entries in $\vs$ and $l$'s are their 
corresponding indices, and the variance $\sigma^2_{ik}$ is given by

\vspace{-4mm}
{\small
\begin{eqnarray}
\sigma^2_{ik}=&& \hspace{-6mm} \text{Var}\bigg(\sum_{j=1, j\neq k}^K{\bf h}_{i,[j]}\vx_j+n_i\bigg) \nonumber\\
&&\hspace{-12mm}=\sum_{j=1\atop{j\neq k}}^K \ \sum_{\vs\in\gsm}
p_{ji}(\vs){\bf h}_{i,[j]} \vs \vs^H {\bf h}_{i,[j]}^H 
- \Big|\hspace{-2mm}\sum_{\vs\in\gsm} \hspace{-2mm} p_{ji}(\vs){\bf h}_{i,[j]}\vs\Big|^2 +\sigma^2\hspace{-1mm}, \nonumber\\ 
&& 
\label{sigma}
\end{eqnarray} }
where $(.)^H$ denotes conjugate transpose operation, and 
$p_{ki}(\vs)$ is the posterior probability given by
\begin{equation}
p_{ki}(\vs) \ \propto \prod_{m=1, m\neq i}^N\exp\Big(\frac{-\big|y_m-\mu_{mk}-{\bf h}_{m,[k]}\vs\big|^2}{2\sigma^2_{mk}}\Big).
\label{pki}
\end{equation}

{\em Message passing:}
The messages exchanged between observation and variable nodes are illustrated 
in Figs. \ref{obs} and \ref{var}. The message from observation node $y_i$ 
to variable node $\vx_k$ consists of the two scalar variables $\mu_{ik}$ and 
$\sigma^2_{ik}$. The message from variable node $\vx_k$ to observation node 
$y_i$ is a vector message given by 
${\bf p}_{ki}=[p_{ki}(\vs_1), p_{ki}(\vs_2),\cdots,p_{ki}(\vs_{|\gsm|})]$. 
The message passing steps are as follows.

\hspace{-4.5mm}
{\bfseries {\em Step 1}}: Initialize $p_{ki}(\vs)$ to $1/|\gsm|$ for all 
$i$, $k$ and $\vs$.\\
{\bfseries {\em Step 2}}: Compute $\mu_{ik}$ and $\sigma^2_{ik}$ from 
(\ref{mu}) and (\ref{sigma}), respectively.\\
{\bfseries {\em Step 3}}: Compute $p_{ki}$ from (\ref{pki}). To improve 
the convergence rate, damping{\footnote{Damping 
can be thought of as reweighting the messages 
with a reweighting parameter (damping factor) $\delta$ 
\cite{damp}-\cite{damp3}.}} of the messages in (\ref{pki}) is done 
with a damping factor $\delta\in (0,1]$,
as shown in the algorithm listing in {\bf Algorithm 1}.
The symbol probabilities at the end of an iteration are computed as 
\begin{eqnarray}
\hspace{-7mm}
p_{k}(\vs) & \hspace{-2.5mm} \propto & \hspace{-3.0mm} \prod_{i=1}^N\hspace{-0.5mm}\exp\hspace{-1mm}\Big(
\frac{-\big|y_i-\mu_{ik}-{\bf h}_{i,[k]}\vs\big|^2}{2\sigma^2_{ik}}\Big),
k=1,\cdots,K.
\label{probs}
\end{eqnarray}

{\em Stopping criterion}: Repeat {\bfseries {\em Steps 2}} 
and {\bfseries {\em 3}} until 
$\|\vp^{(t)}-\vp^{(t+1)}\|<\epsilon$ or until the number 
of iterations is less than a fixed maximum number, where 
{\small
$\vp^{(t)}=[p_1^{(t)}(s_1), \cdots, p_1^{(t)}(s_{|\gsm|}),\cdots\cdots,p_K^{(t)}(s_1), \cdots, p_K^{(t)}(s_{|\gsm|})]^T$
}
and $\epsilon$ is a pre-fixed constant such that $0<\epsilon\ll1$. 

The detected vector of the $k$th user at the BS is obtained as
\begin{eqnarray}
{\hat \vx}_k & \hspace{-2mm} = & \hspace{-2mm} \argmax_{\vs \in \gsm} \ p_{k}(\vs).
\label{argprobs}
\end{eqnarray}
The non-zero entries in ${\hat \vx}_k$ and their indices are then demapped to 
obtain the information bits of the $k$th user. The MP-GSM algorithm listing 
is given in {\bf Algorithm \ref{mpsm}}.

{\em Complexity and performance of MP-GSM algorithm}:
Here, we present and discuss the complexity and
performance of the MP-GSM detection algorithm.

{\em Complexity:}
From (\ref{mu}), (\ref{sigma}), and (\ref{pki}), we see that the total 
complexity of the MP-GSM algorithm is $O(NK|\gsm|)$. 
We compare this complexity with that of the detection using the 
MMSE estimate given by $(\mh^H\mh+\frac{1}{SNR}\mi)^{-1}\mh^H\vy$.
The complexity of this MMSE detection for the system model in (\ref{sysmodel}) 
is given by $O(N^2Kn_t)$. In Table \ref{tab_cmplx}, we present a complexity 
comparison between the MP-GSM and MMSE detection algorithms. From Table
\ref{tab_cmplx}, it can be seen that the MP-GSM detection complexity is 
less than the MMSE detection complexity. In addition to having this 
complexity advantage over MMSE detection, MP-GSM detection achieves 
significantly better performance than MMSE detection (we will see this 
in the performance results presented next). We further note that the 
computation of double summation in (\ref{mu}) and (\ref{sigma}) in the 
MP-GSM algorithm can further be simplified by using FFT, as the double 
summation can be viewed as a convolution operation.

\setlength{\textfloatsep}{2mm}
\begin{algorithm}[t]
{\small
\KwIn{${\bf y}$, ${\bf H}$, $\sigma^2$}
{\bf Initialize}: $p_{ki}^{(0)}(\vs)\gets1/|\gsm|$, $\forall i,k,\vs$

\For{$t = 1 \to {\textit number\_of\_iterations}$ }{
\For{$i = 1 \to N$ }{
\For{$j = 1 \to K$ }{
$\tilde\mu_{ij}\gets$  
$\sum\limits_{\vs\in\gsm}p_{ji}^{(t-1)}(\vs) \sum\limits_{l\in{\cal I}(\vs)}s_{l} H_{i,(j-1)n_t+l}$

$\tilde\sigma^2_{ij}\gets$  
$\scriptstyle{ \sum\limits_{\vs\in\gsm}
p_{ji}^{(t-1)}(\vs){\bf h}_{i,[j]} \vs \vs^H {\bf h}_{i,[j]}^H -|\tilde{\mu}_{ij}|^2}$
}

$\mu_{i}\gets\sum\limits_{j=1}^{K} \tilde\mu_{ij}$

$\sigma_{i}^2\gets\sum\limits_{j=1}^K\tilde\sigma_{ij}^2+\sigma^2$

\For{$k = 1 \to K$ }{
$\mu_{ik}\gets\mu_{i}-\tilde\mu_{ik}$ 

$\sigma^2_{ik}\gets\sigma^2_{i}-\tilde\sigma^2_{ik}$ 

}
}

\For{$k = 1 \to K$ }{
\ForEach{$\vs \in \gsm$}{
$\ln(p_{k}^{(t)}(\vs))\gets\scriptstyle{\sum\limits_{i=1}^N
\frac{-\big|y_i-\mu_{ik}-{\bf h}_{i,[k]}\vs\big|^2}{2\sigma^2_{ik}}}$
}
\For{$i = 1 \to N$ }{
\ForEach{$\vs \in \gsm$}{
$\tilde p_{ki}^{(t)}(\vs)\gets\scriptstyle{ \ln(p_{k}^{(t)}(\vs))+
\frac{\big|y_i-\mu_{ik}-
{\bf h}_{i,[k]}\vs\big|^2}{2\sigma^2_{ik}}}$

$p_{ki}^{(t)}(\vs)=\frac{1-\delta}{C_{ki}}\exp(\tilde p_{ki}^{(t)}(\vs)) + 
\delta p_{ki}^{(t-1)}(\vs)$

$C_{ki}$ is a normalizing constant
}

}
}

}
\KwOut{
$p_{k}(\vs)$ as per (\ref{probs}) and ${\hat \vx}_k$ as per (\ref{argprobs}), 
$\forall k$
}
\label{mpsm}
\vspace{2mm}
\caption{Listing of MP-GSM algorithm. \vspace{2mm}}
}
\end{algorithm}

\begin{figure}
\hspace{-4mm}
\includegraphics[width=3.75in,height=2.75in]{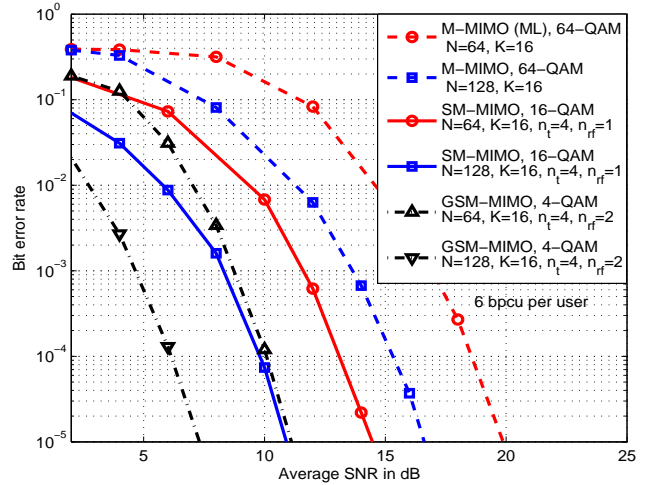}
\vspace{-8mm}
\caption{BER performance of three different multiuser systems with the 
same spectral efficiency of 6 bpcu per user, $K=16$, $N=64,128$: $i$) 
M-MIMO with $n_t=n_{rf}=1$, 64-QAM, sphere decoding; $ii$) SM-MIMO 
with $n_t=4$, $n_{rf}=1$, 16-QAM, MP-GSM detection; $iii$) GSM-MIMO with 
$n_t=4$, $n_{rf}=2$, 4-QAM, MP-GSM detection.}
\vspace{2mm}
\label{mpd6bpcu}
\end{figure}

{\em Performance:}
We evaluated the BER performance of the MP-GSM detection algorithm in
large-scale multiuser GSM-MIMO systems by simulations. Here, we assume 
perfect channel state information (CSI) at the receiver. We will relax 
this assumption later. Figure \ref{mpd6bpcu} presents the performance of 
MP-GSM detection algorithm in a large-scale multiuser GSM-MIMO system 
with the following system parameters:
$K=16$, $N=64,128$, $n_t=4$, $n_{rf}=2$, and 4-QAM. Note that the 
spectral efficiency in this system is 6 bpcu per user. We compare the 
performance of this system with two other systems which also have the 
same spectral efficiency of 6 bpcu per user. These systems are:  
1) conventional multiuser MIMO system with $n_t=n_{rf}=1$, 64-QAM, and
ML detection using sphere decoding (note that this is massive MIMO 
system; we abbreviate it as M-MIMO in the figures), and 2) multiuser 
SM-MIMO system with $n_t=4$, $n_{rf}=1$, 16-QAM, and MP-GSM detection. 
From Fig. \ref{mpd6bpcu}, we observe that GSM-MIMO outperforms 
both SM-MIMO as well as conventional MIMO. For example, at a BER of 
$10^{-3}$, GSM-MIMO has a performance advantage of about 4 dB over 
SM-MIMO and about 7 dB over conventional MIMO. This observation is in 
conformance with similar performance advantage of GSM-MIMO over SM-MIMO 
and conventional MIMO predicted by analytical upper bounds in Section 
\ref{sec3}.

\begin{figure}[h]
\hspace{-4mm}
\includegraphics[width=3.75in,height=2.75in]{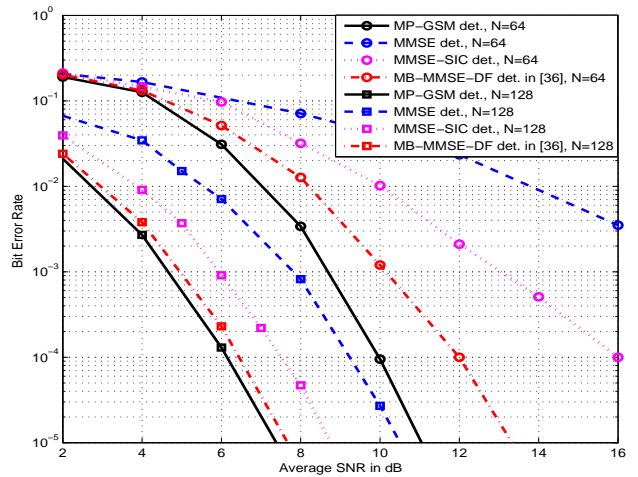} 
\vspace{-10mm}
\caption{
BER performance comparison between $i)$ MP-GSM detector, $ii)$ MMSE 
detector, $iii)$ MMSE-SIC detector, and $iv)$ MB-MMSE-DF 
detector in \cite{lamare}, in multiuser GSM-MIMO with $K=16$, 
$N=64,128$, $n_t=4$, $n_{rf}=2$, 4-QAM, and 6 bpcu per user.
}
\label{mpdvsmmse}
\vspace{-1mm}
\end{figure}

{\em Performance comparison with other detectors:}
Next, in Fig. \ref{mpdvsmmse}, we compare the performance of 
MP-GSM detection with that of MMSE detection in
multiuser GSM-MIMO with $K=16$, $N=64,128$, $n_t=4$, $n_{rf}=2$,
4-QAM, and 6 bpcu per user. From Fig. \ref{mpdvsmmse}, we 
observe that the performance of MP-GSM algorithm is better than MMSE
detection performance by 9 dB for $N=64$ and 3 dB for $N=128$ at 
a BER of $10^{-3}$. As noted in the discussion on complexity,
MP-GSM achieves this better performance than MMSE at a lesser
complexity than MMSE (as illustrated in Table \ref{tab_cmplx}).
Iterative detection/decoding schemes that use MMSE filters 
and provide further refinements are common in the literature
\cite{idd1}-\cite{lamare}. In addition to the comparison with
MMSE detection performance, Fig. \ref{mpdvsmmse} also presents 
a comparison with the performance of the MMSE-SIC detector and
the multi-branch MMSE decision feedback (MB-MMSE-DF) detector 
(with 4 branches and ordering) proposed in \cite{lamare}. It can 
be seen that while the MB-MMSE-DF and MMSE-SIC detectors perform 
better than the MMSE detector, the proposed MP-GSM detector 
outperforms the MMSE, MMSE-SIC, and MB-MMSE-DF detectors. 
Moreover, the complexity of the MP-GSM detector is less than
those of the MMSE, MMSE-SIC, MB-MMSE-DF detectors. 

\begin{table}[h]
\centering
\begin{tabular}{|c||c|c|c|}
\hline
&\multicolumn{3}{|c|}{Complexity in number of real operations 
$\times 10^{6}$}\\
&\multicolumn{3}{|c|}{(GSM-MIMO with $N=128$, $n_t=4, n_{rf}=2$, 4-QAM)}\\
\cline{2-4} $K$ & MMSE & MP-GSM  & CHEMP-GSM\\
& & (Sec. \ref{subsec4a}) & (Sec. \ref{subsec4b}) \\
\hline\hline
16&        3.594&  2.195 & 3.142 \\ \hline
32&       19.767&  4.391 & 6.281 \\ \hline
64&       28.355&  8.782 & 12.265 \\ \hline
96&       36.941& 13.173 & 18.013 \\ \hline
128&      45.526& 17.564 & 21.637 \\ \hline
\end{tabular}
\caption{Comparison between the complexities (in number of real operations)
of MMSE detection, MP-GSM detection, and CHEMP-GSM detection, in multiuser 
GSM MIMO with $N=128$, $n_t=4, n_{rf}=2$ and 4-QAM.}
\vspace{-8mm}
\label{tab_cmplx}
\end{table}

\begin{figure}
\hspace{-4mm}
\includegraphics[width=3.75in,height=2.75in]{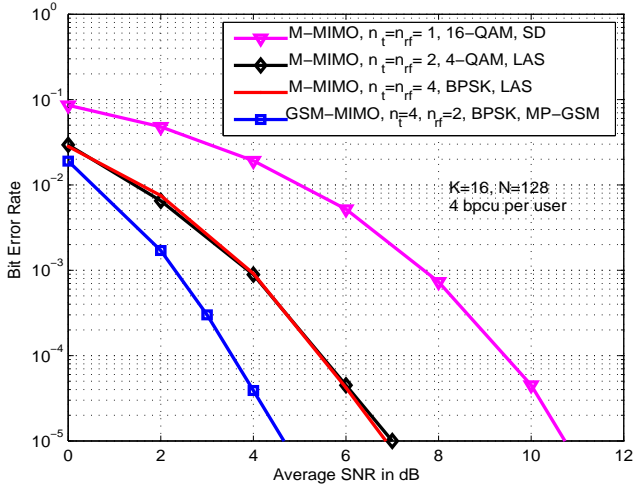}
\vspace{-8mm}
\caption{BER performance of GSM-MIMO with ($n_t=4$, $n_{rf}=2$, BPSK), 
M-MIMO with ($n_t=1$, $n_{rf}=1$, 16-QAM), M-MIMO with 
($n_t=2$, $n_{rf}=2$, 4-QAM), and M-MIMO with ($n_t=4$, $n_{rf}=4$, BPSK) 
for $K=16$, $N=128$, 4 bpcu per user. }
\label{new_fig} 
\vspace{4mm}
\end{figure}

{\em Performance for same spectral efficiency and QAM size:}
We note that if both spectral efficiency and QAM size are to be kept
same in GSM-MIMO and M-MIMO, then the number of spatial streams
per user in M-MIMO has to increase. For example, GSM-MIMO can 
achieve 4 bpcu per user with 4-QAM using $n_t=4$ and $n_{rf}=1$. 
M-MIMO can achieve the same spectral efficiency of 4 bpcu per
user using one spatial stream (i.e., $n_t=1$, $n_{rf}=1$) with 16-QAM. 
But to achieve the same spectral efficiency using 4-QAM in M-MIMO, we 
have to use $n_t=2$, $n_{rf}=2$, i.e., two spatial streams per
user with 4-QAM on each stream are needed. This increase in number 
of spatial streams per user increases the spatial interference.
The effect of increase in number of spatial streams per user
in M-MIMO for the same spectral efficiency on the performance 
is illustrated in Fig. \ref{new_fig} for $K=16$ and $N=128$. 

In Fig. \ref{new_fig}, we compare the performance of the following 
four systems with the same spectral efficiency of 4 bpcu per user,
$K=16$, and $N=128$: 
$i$) GSM-MIMO with ($n_t=4$, $n_{rf}=2$, BPSK), $ii$) M-MIMO with 
($n_t=1$, $n_{rf}=1$, 16-QAM), $iii$) M-MIMO with ($n_t=2$, $n_{rf}=2$, 
4-QAM), and $iv$) M-MIMO with ($n_t=4$, $n_{rf}=4$, BPSK). Detection in 
the GSM-MIMO system is done using the MP-GSM algorithm.
Detection in the M-MIMO systems is done using the likelihood ascent
search (LAS) algorithm in \cite{las} initialized with MMSE solution. 
It can be seen that among the four systems considered in Fig. 
\ref{new_fig}, GSM-MIMO performs the best. This is because M-MIMO 
loses performance because of higher-order QAM or increased spatial 
interference from increased number of spatial streams per user. 

\begin{figure}
\hspace{-4mm}
\includegraphics[width=3.75in,height=2.75in]{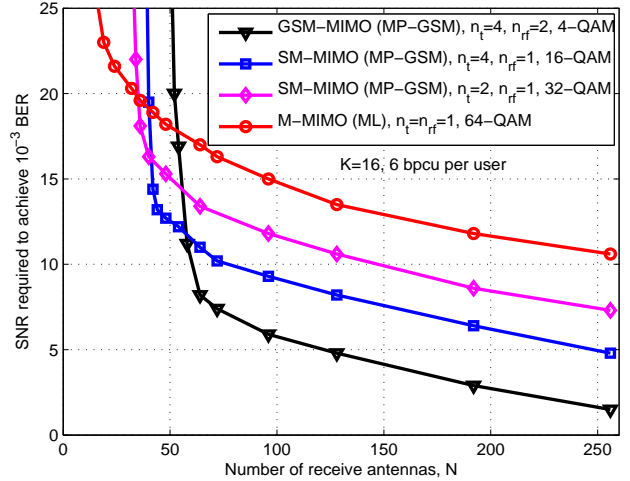} 
\vspace{-8mm}
\caption{Comparison of SNRs required by GSM-MIMO with ($n_t=4$, $n_{rf}=2$, 
4-QAM), SM-MIMO with ($n_t=4$, $n_{rf}=1$, 16-QAM), SM-MIMO with 
($n_t=2$, $n_{rf}=1$, 32-QAM), and M-MIMO with ($n_t=1$, $n_{rf}=1$, 
64-QAM) to achieve a target BER of $10^{-3}$ for $K=16$ and varying 
$N$, at the same spectral efficiency of 6 bpcu per user. }
\label{vary_N} 
\vspace{4mm}
\end{figure}

In Fig. \ref{vary_N}, we fix the number of users at $K=16$ and the
spectral efficiency at 6 bpcu per user, vary the number of antennas
$N$ at the BS, and compare the SNRs required in various systems
to achieve a target BER of $10^{-3}$. We compare the performance of the 
following four systems: $i$) GSM-MIMO with ($n_t=4$, $n_{rf}=2$, 4-QAM), 
$ii$) SM-MIMO with ($n_t=4$, $n_{rf}=1$, 16-QAM), $iii$) SM-MIMO with 
($n_t=2$, $n_{rf}=1$, 32-QAM), and $iv$) M-MIMO with ($n_t=1$, $n_{rf}=1$, 
64-QAM). From Fig. \ref{vary_N}, it can be observed that as the number of 
antennas at the BS increases, the required SNR to achieve the target BER 
decreases in all the four systems, which is expected because of the 
increased receive diversity. The sharp degradation observed for small 
values of $N$ is because the systems become under-determined when 
$Kn_t > N$, and hence the required SNRs shoot up. When $Kn_t\leq N$ 
(fully/over-determined), GSM MIMO outperforms M-MIMO by about 9 dB 
and SM-MIMO by about 4 to 6 dB. 

\subsection{CHEMP-GSM detection algorithm}
\label{subsec4b}
In this subsection, we propose another detection algorithm based on message
passing. We refer to the detection algorithm presented in this subsection 
as CHEMP-GSM (channel hardening exploiting message passing 
\cite{chemp},\cite{chemp_sm} for GSM) 
algorithm. We also propose a novel channel estimator for use in the 
CHEMP-GSM detector. We refer to the CHEMP-GSM detector along with this 
channel estimator as the `CHEMP-GSM receiver'. 
CHEMP-GSM approach is another message passing approach which gives
less complexity than MMSE but performs significantly better than MMSE.

{\em Matched filtered system model:}
First, we perform a matched filter operation on the received signal 
vector $\vy$ in (\ref{sysmodel}) as $\mh^H\vy$, which can be written as
\begin{eqnarray}
\mh^H\vy = \mh^H(\mh\vx + \vn).
\label{mfeq}
\end{eqnarray}
A corresponding equivalent system model can be written as
\begin{equation}
\vz =\mj\vx+\vv,
\label{eqn1}
\vspace{-4mm}
\end{equation}
\vspace{-2mm}
where
\begin{equation}
\vz \Define \frac{\mh^H\vy}{N}, \quad
\mj\Define\frac{\mh^H\mh}{N}, \quad \vv\Define\frac{\mh^H\vn}{N}.
\label{def1}
\end{equation}
Similar to $\vx$ in the basic system model in (\ref{sysmodel}), the vector 
$\vz$ in (\ref{eqn1}) can be viewed as a concatenation of $K$ sub-vectors 
each of dimension $n_t\times1$, i.e., 
$\vz = [\vz_1^T \ \ \vz_2^T\,\cdots\,\vz_k^T\,\cdots\,\vz_K^T]^T$.
Likewise, $\vv = [\vv_1^T \ \ \vv_2^T\,\cdots\,\vv_k^T\,\cdots\,\vv_K^T]^T$,
where $v_j=\sum_{i=1}^{N}\frac{H_{ij}^*n_i}{N}$ is the $j$th element 
of $\vv$ and $H_{ij}$ is the $(i,j)$th element of ${\bf H}$.  For large $N$, 
$v_j$ can be approximated to follow Gaussian distribution as 
$v_j\sim {\cal CN}(0,\sigma^2_{v})$, where the variance 
$\sigma^2_{v}=\frac{\sigma^2}{N}$. Each sub-vector 
$\vz_k$ can be expressed as
\begin{equation}
\vz_k \ = \ \mj_{kk}\vx_k+\underbrace{\sum_{j=1, j\neq k}^{K} \mj_{kj}\vx_j + \vv_k}_{\Define \ \vg_k},
\label{eqn2}
\end{equation}
where $\mj_{kj}$ is a $n_t\times n_t$ sub-matrix of $\mj$ formed from the 
elements in rows $(k-1)n_t+1$ to $kn_t$ and columns $(j-1)n_t+1$ to $jn_t$,
i.e., $\mj$ can be written in terms of the sub-matrices as 
\[
\mj= \begin{bmatrix} 
\mj_{11} & \mj_{12} & \cdots & \mj_{1K} \\
\mj_{21}   & \mj_{22} & \cdots & \mj_{2K} \\
\vdots & & \ddots & \vdots \\
\mj_{K1} & \mj_{K2} & \cdots & \mj_{KK}
\end{bmatrix}. \] 
The vector $\vg_k$ defined in (\ref{eqn2}) denotes the interference-plus-noise 
to the $k$th user's GSM signal. This term $\vg_k$ involves the off-diagonal 
elements of ${\bf J}$ (i.e., $J_{kj}$, $k\neq j$ where $J_{kj}$ 
is the ($k,j$)th element in ${\bf J}$). Due to channel
hardening that occurs in large MIMO channels, the matrix $\mj$ (and hence 
$\mj_{kk}, \forall k$) has strong diagonal elements compared to off-diagonal
terms for large $N,K$. We approximate $\vg_k$ to be multivariate Gaussian 
with mean $\vmu_k$ and covariance $\msig_k$, which can be written as 
\begin{eqnarray}
\vmu_k& \hspace{-2mm}=& \hspace{-2mm} \E(\vg_k)=\sum_{j=1, j\neq k}^{K} \hspace{-2mm} \mj_{kj} \E(\vx_j) \label{eq_mu} \\
\msig_k& \hspace{-2mm}=& \hspace{-2mm}\text{Cov}(\vg_k)=\sum_{j=1, j\neq k}^{K} \hspace{-2mm} \mj_{kj}\text{Cov}(\vx_j)\mj_{kj}^H + \sigma^2_{v}\mi_{n_t},
\label{eq_sigma}
\vspace{-2mm}
\end{eqnarray}
where
\vspace{-2mm}
\begin{eqnarray}
\E(\vx_j)& \hspace{-2mm} =& \hspace{-2mm}\sum_{\forall \vs, \ \vs \in \gsm}\hspace{-2mm} \vs p_j(\vs) \\
\text{Cov}(\vx_j)&\hspace{-2mm}=&\hspace{-2mm}\sum_{\forall \vs, \ \vs \in \gsm} \hspace{-2mm} \vs \vs^H p_j(\vs)-\E(\vx_j)\E(\vx_j)^H,
\end{eqnarray}
and 
\vspace{-3mm}
\begin{equation}
p_k(\vs)=\Pr\big(\vx_k= \vs\big), \quad \vs \in \gsm.
\end{equation}
Let $\vp_k=[p_{k}(\vs_1), p_{k}(\vs_2),\cdots,p_{k}(\vs_{|\gsm|})]$ denote 
the vector of probability masses corresponding to GSM signal vector $\vx_k$. 
The probability masses $p_k(\vs)$'s are approximated with the corresponding
a posteriori probabilities (APP), i.e., 

\vspace{-3mm}
{\small
\begin{eqnarray}
\hspace{-4mm}p_k(\vs)&\hspace{-3mm}\gets&\hspace{-3mm}\Pr(\vx_k=\vs|\vz_k, \mj) \label{probs_x} \\
&\hspace{-3mm} \propto & \hspace{-3mm}
\exp\Big(\frac{-1}{2}(\vz_k-\mj_{kk}\vs-\vmu_k)^H\msig_k^{-1}(\vz_k-\mj_{kk}\vs-\vmu_k)\Big).
\label{probs_x1}
\end{eqnarray}}

\begin{figure}
\centering
\includegraphics[width=1.75in,height=2.5in]{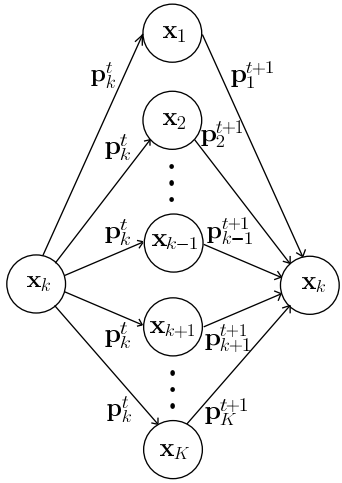}
\caption{Message passing in CHEMP-GSM algorithm.}
\vspace{2mm}
\label{fig_new}
\end{figure}

\vspace{-2mm}
{\em Message passing:}
The system is modeled as a fully-connected graph with $K$ nodes, where 
the $k$th node is an approximate APP processor corresponding to $\vx_k$. 
Node $k$ computes the APP based on the incoming messages and the knowledge 
of $\vz_k$ and $\mj$. The incoming messages to node $k$ are the APPs from 
the remaining nodes. The APP computed in node $k$, in turn, is passed to 
the remaining nodes for their APP computation in the next iteration. The 
messages exchanged between the nodes are illustrated in Fig. \ref{fig_new}. 
The algorithm is as follows.

\hspace{-5.5mm}
{\bfseries \em Step 1:} 
The probability vectors $\vp_k$'s are initialized with equiprobable masses 
$1/|\gsm|$.  \\
{\bfseries \em Step 2:} 
Node $k$ computes $\vp_k$ as per (\ref{probs_x1}) with the incoming 
vector messages
{\small $\{\vp_1,\vp_2,\cdots,\vp_{k-1},\vp_{k+1},\cdots,\vp_{K}\}$}. 
Damping of messages with damping factor $\delta\in [0,1)$ is done to 
improve the rate of convergence. That is, if $\tilde{\vp}_k^t$ is the 
computed probability vector at the $t$th iteration, the message at the 
end of $t$th iteration is computed as 
\begin{equation}
\vp_k^{t} \ = \ (1-\delta)\tilde{\vp}_k^t+\delta \vp_k^{t-1}.
\label{damp}
\end{equation}
Repeat {\bfseries \em Step 2} for a certain number of iterations, 
after which the algorithm stops. The estimate of the $l$th modulation 
symbol transmitted by the $k$th user is obtained as 
\vspace{-1mm}
\begin{equation}
{\hat s}_{k,l} \ = \ \argmax_{s \in \sa} \ \sum_{\forall \vs, \ \vs \in \gsm \, : \, {\cal X}_l(\vs)=s}p_k(\vs),
\label{modbits}
\end{equation}
where ${\cal X}_l(\vs)$ is the $l$th non-zero element in $\vs$ and 
$l\in\{1,2,\cdots,n_{rf}\}$. An estimate of the active antenna 
indices chosen for transmission by the $k$th user is obtained as
\begin{equation}
{\hat q}_k \ = \argmax_{q \in \{1,\cdots, |\sap|\}} \ \sum_{\forall \vs, \ \vs \in \gsm \, : \, {\cal I}(\vs)=q}p_k(\vs).
\label{indexbits}
\end{equation}
The values of ${\hat s}_{k,l}$ and ${\hat q}_k$ are then demapped to obtain the 
information bits of the $k$th user. 

{\em Complexity}:
The orders of complexity for the computation of $\vz$ and $\mj$ are 
$O(NKn_t)$ and $O(NK^2n_t^2)$, respectively. The complexities for the 
computation of (\ref{eq_mu}), (\ref{eq_sigma}) and (\ref{probs_x1}) are 
of orders $O(n_t^2K^2)$, $O(n_t^3K^2)$ and $O(n_t^3K|\gsm|)$, respectively. 
For $N>K$ and $K>n_t$, the overall complexity of the algorithm is dominated 
by the computation of $\mj$ whose complexity is $O(NK^2n_t^2)$. 
Therefore, the overall complexity of the algorithm is $O(NK^2n_t^2)$. 
This complexity is less than the MMSE detection complexity which is 
$O(N^2Kn_t)$. This is illustrated numerically in Table \ref{tab_cmplx}  
which shows the complexity comparison between MMSE and CHEMP-GSM detection.

\begin{figure}
\hspace{-4mm}
\includegraphics[width=3.75in,height=2.85in]{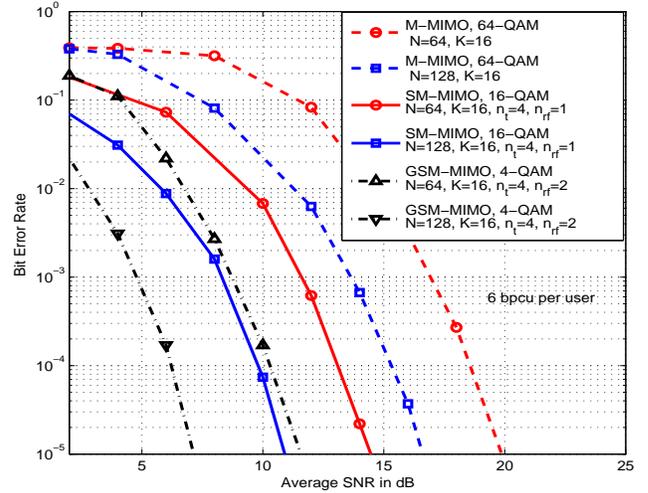}
\vspace{-8mm}
\caption{BER performance of multiuser GSM-MIMO systems with 
($n_t=4$, $n_{rf}=2$, 4-QAM) and  ($n_t=4$, $n_{rf}=1$ using CHEMP-GSM 
detection, and M-MIMO system with ($n_t=1$, $n_{rf}=1$, 64-QAM) using 
sphere decoding, at 4 bpcu per user, $K=16$, $N=64,128$.}
\label{cdt6bpcu}
\end{figure}

\begin{figure}[h]
\hspace{-4mm}
\includegraphics[width=3.75in,height=2.85in]{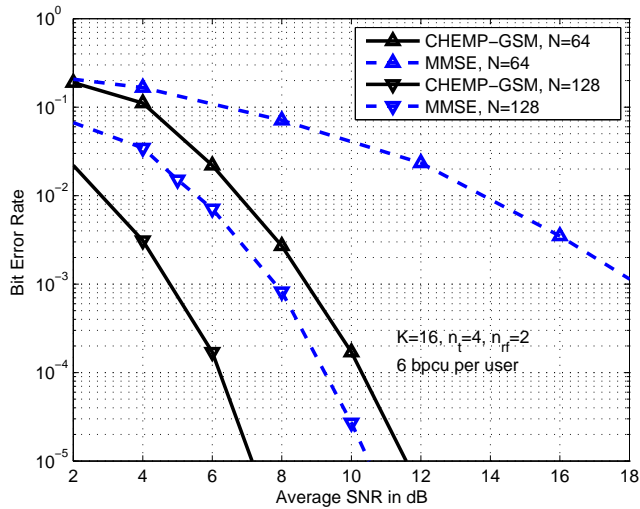}
\vspace{-8mm}
\caption{BER performance comparison of multiuser GSM-MIMO using CHEMP-GSM 
detector and MMSE detector for $K=16, n_t=4$, $n_{rf}=2$, 4-QAM at 4 
bpcu per user, $N=64,128$. }
\label{cdtvsmmse}
\vspace{2mm}
\end{figure}

{\em Performance:} 
In Fig. \ref{cdt6bpcu}, we present the performance of the 
CHEMP-GSM detector at a spectral efficiency of 6 bpcu per user, $K=16$, 
$N=64,128$, assuming perfect CSI at the receiver. Figure \ref{cdt6bpcu} 
compares the performance of multiuser GSM-MIMO system with 
($n_t=4$, $n_{rf}=2$, 4-QAM), with that of SM-MIMO system with 
($n_t=4$, $n_{rf}=1$, 16-QAM) and M-MIMO system with ($n_t=1$, $n_{rf}=1$, 
64-QAM). Detection in GSM-MIMO and SM-MIMO systems is done using CHEMP-GSM 
algorithm. Detection in M-MIMO system is done using sphere decoding. As 
observed in Sec. \ref{subsec_new} and Sec. \ref{subsec4a}, in Fig. 
\ref{cdt6bpcu} also we see that GSM-MIMO outperforms both SM-MIMO and 
M-MIMO. This is because of the smaller-sized QAM used in GSM-MIMO 
compared to those used in SM-MIMO and M-MIMO to achieve the same 
spectral efficiency. 

{\em CHEMP-GSM vs MMSE performance:}
Figure \ref{cdtvsmmse} shows the performance comparison between CHEMP-GSM 
detector and MMSE detector at a spectral efficiency of 6 bpcu per user, 
$K=16, n_t=4, n_{rf}=2$, and 4-QAM. It is observed that CHEMP-GSM detector 
outperforms MMSE detector by about 9 dB for $N=64$ and 3 dB for $N=128$ at 
a BER of $10^{-3}$. We note that CHEMP-GSM detector achieves this better 
performance at a lesser complexity compared to MMSE detector. This can be 
observed from Table \ref{tab_cmplx}.

\begin{figure}
\hspace{-4mm}
\includegraphics[width=3.75in,height=2.85in]{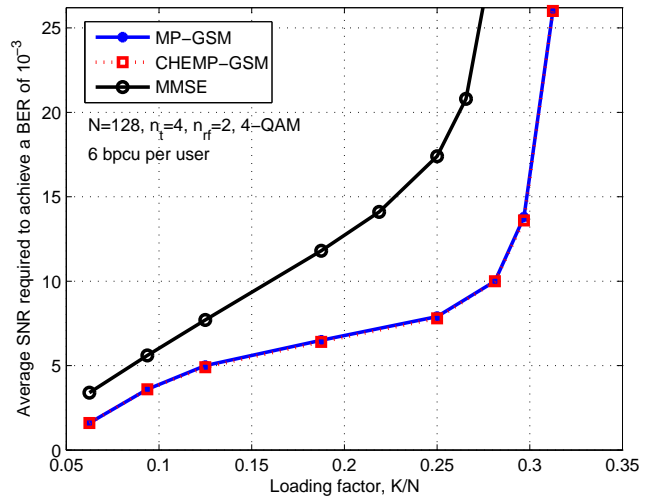}
\vspace{-8mm}
\caption{SNR required by MP-GSM, CHEMP-GSM, and MMSE detectors to achieve 
a BER of $10^{-3}$ with $N=128, n_t=4$, $n_{rf}=2$, 4-QAM, and 6 bpcu per 
user, as a function of system loading factor $K/N$. }
\label{alpha_p} 
\vspace{2mm}
\end{figure}

{\em Performance as a function of loading factor:}
In Fig. \ref{alpha_p}, we compare the performance of different
detectors for multiuser GSM-MIMO with $N=128, n_t=4$, $n_{rf}=2$, 
4-QAM, and 6 bpcu per user as a function of system loading factor $K/N$.
We plot the average SNR required by MP-GSM, CHEMP-GSM and MMSE detectors
with perfect CSI to achieve a BER of $10^{-3}$. From Fig. \ref{alpha_p}, 
we observe that the proposed message passing based detectors outperform 
the MMSE detector by about 2 to 3 dB at lower loading factors. This gap 
widens as the system loading factor increases; for example, the gap is
about 3 dB at $K/N=0.125$, and it widens to about 9 dB at $K/N=0.25$. The
SNR required significantly increases for high loading factors because 
the channel gain matrix which is of dimension $N \times Kn_t$ becomes 
under-determined for $K>\frac{N}{n_t}$. Also, both MP-GSM and CHEMP-GSM
algorithms perform almost same, with MP-GSM having a lesser complexity 
than CHEMP-GSM (see Table \ref{tab_cmplx}). However, the matched filtered
system model in CHEMP-GSM allows a simple channel estimation technique,
which is presented in the following subsection.

\subsection{Estimation of ${\bf H}^H{\bf H}$}
\label{subsec4f}
In obtaining the performance results reported in the previous subsection, 
we have assumed perfect CSI at the receiver. Now, we relax this assumption. 
We present a channel estimation scheme suited for use in CHEMP-GSM detector. 
A conventional approach is to directly obtain an estimated channel matrix 
$\widehat{\mh}$ through channel estimation techniques (MMSE channel 
estimation, for example) using pilot transmissions, and use $\widehat{\mh}$ 
in place of $\mh$ in detection algorithms. For the MMSE detector and MP-GSM 
detector, we follow this approach, and we call the receiver employing MMSE 
detector with MMSE channel estimate as `MMSE receiver' and the receiver 
employing MP-GSM detector with MMSE channel estimate as `MP-GSM receiver'.
For CHEMP detector, however, instead of conventional approaches that 
estimate $\mh$, we directly obtain an estimate of the matrix $\mj$. The 
motivation for this approach is that $\mh$ influences the proposed 
detection operation through $\mj=\mh^H\mh$. 

Assume that the channel is slowly fading, and that the fade remains 
constant for one frame duration. Each frame has $L_f$ channel uses, 
consisting of a pilot part and a data part. The pilot part consists of 
$Kn_t$ channel uses, and the data part consists of $L_f-Kn_t$ channel 
uses. Let $\mx_{\text p}=A\mi_{Kn_t}$ denote the pilot matrix, where, in 
the $j$th channel use, $1\leq j\leq Kn_t$, the $\lceil\frac{j}{n_t}\rceil$th 
user terminal transmits a pilot symbol with amplitude $A$ through its 
antenna whose index is given by $\big((j-1)\mod n_t\big)+1$, and the other 
antennas remain silent. The signal received at the BS during pilot part is 
given by
\begin{eqnarray} 
\my_{\text p} & \hspace{-1mm} = & \hspace{-1mm} \mh\mx_{\text p}+\mw_{\text p} 
\ = \  A\mh+\mw_{\text p},
\end{eqnarray}
where $A=\sqrt{KE_s}$, $E_s$ is the average symbol energy, and 
$\mw_{\text p}$ is 
the noise matrix. An estimate of the matrix $\mj$ is obtained as
\vspace{-4mm} 
\begin{eqnarray}
\mje & \hspace{-1mm} = & \hspace{-1mm} \frac{\my_{\text p}^H\my_{\text p}}{NA^2}-\frac{\sigma^2_v}{A^2}\mi_{Kn_t}.
\label{mpmdj}
\end{eqnarray}
where $\sigma^2_{v}=\frac{\sigma^2}{N}$.
An estimate of the vector $\vz$ is obtained as
\begin{eqnarray}
\vze & \hspace{-1mm} = & \hspace{-1mm} \frac{\my_{\text p}^H\vy}{NA},
\label{mpmdz}
\end{eqnarray}
where $\vy$ is the received signal vector in the data phase. 
These estimates $\mje$ and $\vze$ in (\ref{mpmdj}) and (\ref{mpmdz}) are 
used in the CHEMP-GSM detection algorithm in place of $\mj$ and $\vz$.

\begin{figure}
\hspace{-3mm}
\subfigure[BER vs SNR.]{
\includegraphics[width=1.75in,height=2.35in]{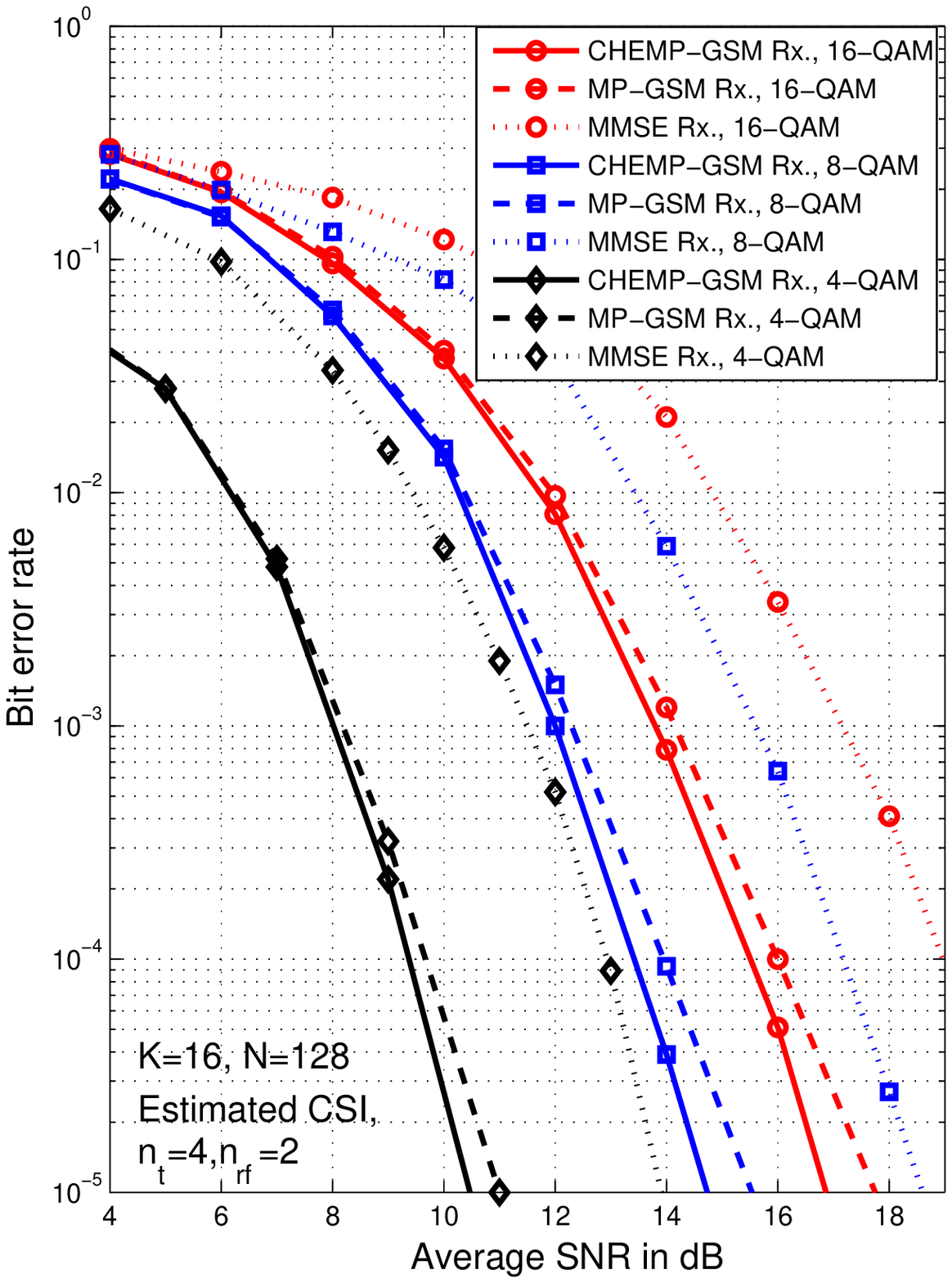}
\label{new_sys1}
}
\hspace{-6mm}
\subfigure[SNR reqd. for $10^{-3}$ BER vs $K/N$.]{ 
\includegraphics[width=1.8in,height=2.35in]{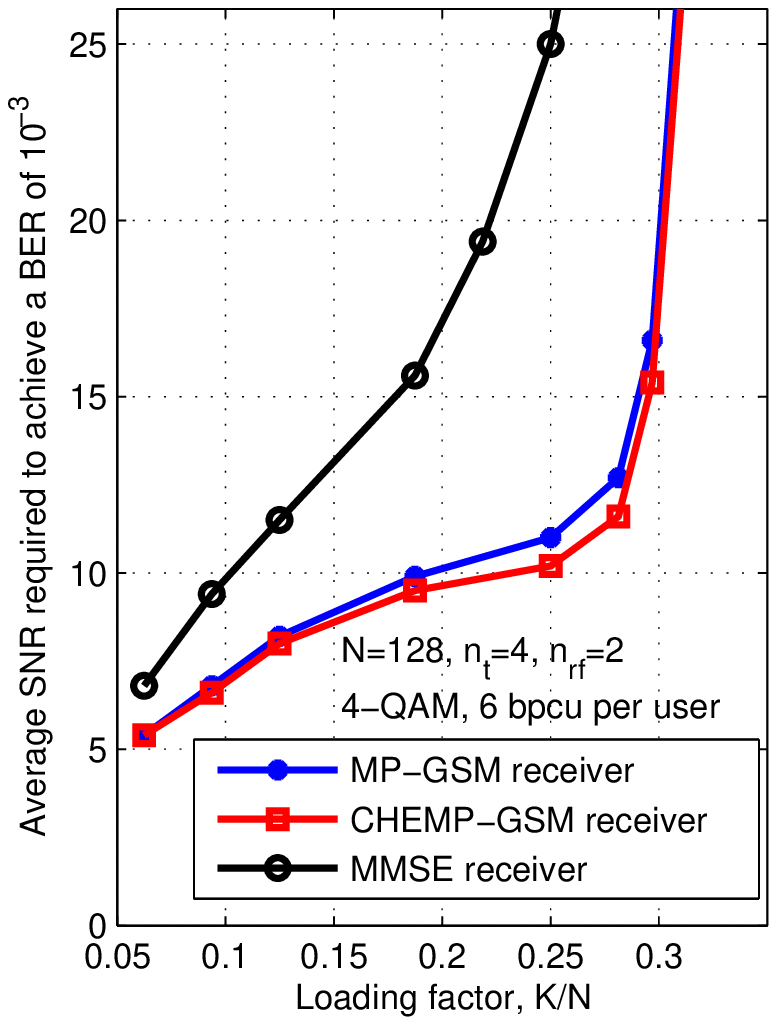}
\label{new_sys2}
}
\caption{Performance of MMSE Rx, MP-GSM Rx, and CHEMP-GSM Rx in multiuser
GSM-MIMO with $N=128$, $n_t=4$, $n_{rf}=2$, 4-/8-/16-QAM,
and estimated CSI.
}
\vspace{4mm}
\end{figure}

{\em Performance and complexity:} 
In Figs. \ref{new_sys1} and \ref{new_sys2}, we present performance 
comparisons between $i$) MMSE receiver (MMSE detector with MMSE channel 
estimate), $ii$) MP-GSM receiver (MP-GSM detector with MMSE channel 
estimate), and $iii$) CHEMP-GSM receiver (CHEMP-GSM detector with the
proposed estimate of $\mj$). Multiuser GSM-MIMO with $N=128$, $n_t=4$, 
$n_{rf}=2$, and 4-/8-/16-QAM is considered. Figure 
\ref{new_sys1} shows the BER vs SNR plots for $K=16$, i.e., loading 
factor $K/N=0.125$. Figure \ref{new_sys2} shows SNR required to achieve 
$10^{-3}$ BER as a function of loading factor $K/N$. Table \ref{comp2} 
presents corresponding complexity comparison between the receivers 
considered. It can be observed that, because of the additional complexity 
of MMSE channel estimation, the complexities of MMSE receiver and MP-GSM
receiver are more than the corresponding complexities of MMSE detector 
and MP-GSM detector in Table \ref{tab_cmplx}. However, the complexities 
of CHEMP-GSM receiver and CHEMP-GSM detector are the same. This is because
the complexities of computing $\mje$ and $\vze$ are the same as those of
computing $\mj$ and $\vz$, respectively. 
From Figs. \ref{new_sys1} and \ref{new_sys2}, it can be seen 
that the performance of the CHEMP-GSM receiver is the best among the three 
receivers considered. Also, the complexities of both MP-GSM and CHEMP-GSM 
receivers are much less compared to that of MMSE receiver, and the complexity 
of CHEMP-GSM receiver is a little higher than the complexity of MP-GSM 
receiver due to the additional computation of ${\vze}$ in (\ref{mpmdz}).
  
\begin{table}[t]
\centering
\begin{tabular}{|c||c|c|c|}
\hline
&\multicolumn{3}{|c|}{Complexity in number of real operations $\times
10^{6}$, $N=128$}\\
\cline{2-4} $K$ & \hspace{4mm}MMSE  Rx. \hspace{4mm}& MP-GSM Rx.&
CHEMP-GSM Rx.\\
\hline\hline
16&        4.041& 2.466 & 3.142 \\ \hline
32&       21.294& 4.925 & 6.281 \\ \hline
64&       30.037& 10.831 & 12.265 \\ \hline
96&       39.513& 16.746 & 18.013 \\ \hline
128&      48.622& 20.961 & 21.637 \\ \hline
\end{tabular}
\caption{Comparison between the complexities (in number of real operations)
of the MP-GSM receiver, CHEMP-GSM receiver, MMSE receiver in GSM-MIMO
with $N=128$, $n_t=4$, $n_{rf}=2$, and 4-QAM.}
\vspace{-2mm}
\label{comp2}
\end{table}

\section{Multiuser GSM-MIMO in frequency selective fading }
\label{sec5}
In this section, we assume the multiuser GSM-MIMO system model described 
in Section \ref{sec2}, except for the channel model which is now assumed 
to be frequency selective. Let $L$ denote the number of multipath 
components between each pair of transmit antenna at the user and receive 
antenna at the BS. Let $H_{i,(k-1)n_t+j}^{(l)}$ denote the channel gain 
from the $j$th transmit antenna of the $k$th user to the $i$th BS receive 
antenna on the $l$th multipath component. The channel gains for the $l$th 
multipath component are modeled as complex Gaussian with zero mean and 
variance $\Omega_l$. The power-delay profile of the channel 
is modeled as 
\begin{eqnarray}
\Omega_l & \hspace{-1mm} = & \hspace{-1mm} \E[|H_{i,(k-1)n_t+j}^{(l)}|^2] \nonumber \\
&\hspace{-1mm} = & \hspace{-1mm} \Omega_{0}10^{-\xi l/10}, \ \ l=0,\cdots,L-1.
\end{eqnarray}
where $\xi$ denotes the decay-rate of the average power in each of the
multipath components in dB. The total power of the channel is assumed to 
be unity, i.e., $\sum_{l=0}^{L-1} {\Omega_l} = 1$. We will further assume 
that the channel coefficients are estimated at the BS using a pilot 
transmission based channel estimation.

\begin{figure}
\centering
\includegraphics[width=3.25in,height=2.0in]{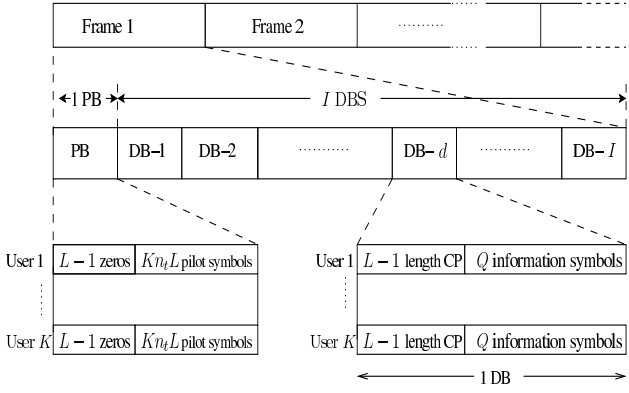} 
\caption{Frame structure of CPSC scheme for multiuser GSM-MIMO in 
frequency selective fading.}
\label{fig_frame} 
\vspace{2mm}
\end{figure}

{\em CPSC transmission}:
We employ cyclic prefixed single carrier (CPSC) transmission, 
which has the advantage of low peak-to-average power ratio (PAPR) 
\cite{cpsc1},\cite{cpsc2}. Transmission is carried out in frames, where 
each frame consists of multiple blocks as shown in Fig. \ref{fig_frame}. 
The fade coefficients are assumed to be constant over one frame duration. 
Each frame consists of one pilot block (PB) meant for channel estimation, 
followed by $I$ data blocks (DB). The pilot block consists of $(L-1)+Kn_tL$ 
channel uses. In the first $L-1$ channel uses in the pilot block, zeros are 
transmitted to avoid interference from previously transmitted frames. In 
each of the remaining $Kn_tL$ channel uses, a $Kn_t$-length pilot symbol 
vector comprising of pilot symbols transmitted from $K$ users ($n_t$ pilot 
symbols per user) is received by the $N$ BS receive antennas. Each data 
block consists of $Q+L-1$ channel uses, where $Q$ information symbol 
vectors, each of length $Kn_t$, prefixed by $(L-1)$-length cyclic prefix 
from each user are transmitted. With $I$ data blocks in a frame, the 
number of channel uses in the data part of the frame is $(Q + L-1)I$. 
Taking both pilot and data channel uses into account, the total number 
of channel uses per frame is $(Kn_t + 1)L + (Q + L-1)I-1$.

{\em Channel estimation:} Let $P$ be the power of the pilot symbol 
transmitted by each user. Then, in $Kn_tL$ channel uses, the pilot
sequence ${\mathbf p}^k_j$ transmitted by the $j$th transmit antenna 
of the $k$th user is given by the $Kn_tL$-length vector
$
{\mathbf p}^k_j=[{\mathbf 0}_{((k-1)n_t+j-1)L\times1} \ \ \sqrt{P} \ \ {\mathbf 0}_{(((K-k+1)n_t-j+1)L-1)\times1}].
$
Let

\vspace{-5mm}
{\small\[
\vh_{i,(k-1)n_t+j}=[H_{i,(k-1)n_t+j}^{(0)}\,\cdots\,H_{i,(k-1)n_t+j}^{(l)}\,\cdots\,H_{i,(k-1)n_t+j}^{(L-1)}].
\]}

\vspace{-5.5mm}
\hspace{-5mm}
The $Kn_tL$-length vector received at the $i$th BS antenna in the pilot
phase is then given by
\begin{eqnarray}
\hspace{-6mm}
\vy^i_{\mbox{\tiny P}}&\hspace{-2mm}=& \hspace{-2mm}[y^{i,0}_{\mbox{\tiny P}}\, \, y^{i,1}_{\mbox{\tiny P}}\,\, \cdots \,\, y^{i,Kn_tL-1}_{\mbox{\tiny P}}]\nonumber\\
&\hspace{-2mm}=& \hspace{-2mm}\underbrace{[\vh_{i,1}^T\,\cdots\,\vh_{i,(k-1)n_t+j}^T\,\cdots\,\vh_{i,Kn_t}^T]^T}_{\Define \ \vh^i} \sqrt{P} + \vn^i_{\mbox{\tiny P}},
\label{pilot}
\end{eqnarray}
where $y^{i,t}_{\mbox{\tiny P}}$ is the received signal at the $i$th BS 
antenna in the $t$th channel use of the pilot phase, and 
$\vn^i_{\mbox{\tiny P}}$ is the $Kn_tL$-length noise sequence at the $i$th 
BS antenna. An MMSE estimate of the channel gain vector $\vh^i$ can be 
obtained from (\ref{pilot}) as
\begin{equation}
\widehat{\vh}^i \ = \ \frac{\sqrt{P}}{P+\sigma^2} \vy^i_{\mbox{\tiny P}}.
\label{est}
\end{equation}

{\em Signal detection:} Each data block (DB) in a frame consists of a 
cyclic prefix (CP) followed by data symbols as shown in Fig. \ref{fig_frame}. 
Let $\vx_k^{(t)} \in \gsm$ denote the transmit vector from the $k$th user in
the $t$th channel use in a DB. The $k$th user's transmit vector in a DB is 
of the form  

\vspace{-4mm}
{\small
\begin{equation}
[\underbrace{ {\vx_k^{(Q-L)}}^T\, {\vx_k^{(Q-L+1)}}^T\, \cdots \, {\vx_k^{(Q-1)}}^T\,}
_{\scriptsize{\mbox{CP}}}\,\underbrace{{\vx_k^{(0)}}^T\, {\vx_k^{(1)}}^T\, 
\cdots \, {\vx_k^{(Q-1)}}^T}_{\scriptsize{\mbox{Data}}}]^T. \nonumber
\end{equation}
}

\vspace{-4mm}
Assuming perfect synchronization and discarding the CP at the BS receiver, 
the received signal vector can be written as 
\begin{equation}
\vy' =  \mh'\vx'+\vn',
\label{cpsc_sys}
\end{equation}
where $\vy'$ is 
$[{\vy^{(0)}}^T \,{\vy^{(1)}}^T\cdots\,{\vy^{(Q-1)}}^T]^T\in \mathbb{C}^{NQ\times 1}$, 
$\vy^{(t)}\in \mathbb{C}^{N\times 1}$ denotes the received vector at the
$t$th channel use in a DB, $\vx'$ is 
$[{\vx^{(0)}}^T \,{\vx^{(1)}}^T,\cdots\,{\vx^{(Q-1)}}^T]^T\in \mathbb{C}^{Kn_tQ\times 1}$,
$\vx^{(t)}\in \mathbb{C}^{Kn_t\times 1}$ is the vector comprising of 
transmit vectors of all the users in the $t$th channel use in a DB, 
$\mh'$ is the channel gain matrix of dimension $NQ\times Kn_tQ$, 
and $\vn'$ is the additive white Gaussian noise vector given by
$[{\vn^{(0)}}^T \,{\vn^{(1)}}^T\cdots\,{\vn^{(Q-1)}}^T]^T \in \mathbb{C}^{NQ\times 1}$. 
The received signal vector in the $t$th channel use in a DB can 
be written as
\begin{equation}
\vy^{(t)} = \sum_{l=0}^{L-1} \mh^{(l)}\vx^{(t-l)} + \vn^{(t)}, \ \ 
t=0,1,\cdots,Q-1,
\end{equation} 
where $\mh^{(l)} \in \mathbb{C}^{N\times Kn_t}$ is the channel gain 
matrix corresponding to the $l$th multipath component such that 
$H_{i,(k-1)n_t+j}^{(l)}$ represents the channel gain from the $j$th 
transmit antenna of the $k$th user to the $i$th BS receive antenna in 
the $l$th multipath. For this system, the ML detection rule is given by
\begin{equation} 
\label{mlq} 
{\hat{\vx}}' \ = \ \argmin_{{\vx'}\in \gsmk^{Q}} 
\|{\vy'}-{\mh'}{\vx'}\|^2,
\end{equation}
where $\gsmk\Define(\gsm)^K$, and the exact computation of (\ref{mlq}) 
requires exponential complexity in $KQ$. We shall formulate the system 
model in (\ref{cpsc_sys}) into an equivalent system model in the 
frequency domain, and employ the algorithms for signal detection 
(presented in Section \ref{sec4}) on the resulting equivalent 
system model. 

It is noted that because of the addition of CP, the matrix $\mh'$ is a 
block circulant matrix. Therefore, $\mh'$ can be transformed into a block 
diagonal matrix ${\mathbf D}$ as
\begin{equation} 
{\mathbf D} \ = \ (\mathbf{F} \otimes \mi_{N})\mh'(\mathbf{F}^H \otimes \mi_{Kn_t}),
\end{equation}
where $\mi_n$ denotes $n\times n$ identity matrix, and  ${\mathbf F}$ is 
the $Q\times Q$ DFT matrix, given by
\begin{equation}
 {\mathbf F} = \frac{1}{\sqrt{Q}}\begin{bmatrix}
\rho_{0,0} &\rho_{0,1} & \cdots  & \rho_{0,Q-1} \\
\rho_{1,0} & \rho_{1,1} & \cdots & \rho_{1,Q-1} \\
  \vdots  & \vdots  & \vdots & \vdots  \\
  \rho_{Q-1,0} & \rho_{Q-1,1} & \cdots & \rho_{Q-1,Q-1} \end{bmatrix},
\nonumber
\end{equation}  
where $\rho_{u,v} = \exp(-{\bf j}\frac{2\pi uv}{Q})$. $\mathbf{D}$ is a block 
diagonal matrix of the form
\begin{equation}
\mathbf{D} = \begin{bmatrix}
{\mathbf{D}_0}  & \cdots  & 0 \\
 \vdots & \ddots & \vdots \\
  0  & \cdots & \mathbf{D}_{Q-1} \end{bmatrix}, 
\end{equation}
where $\mathbf{D}_q$ is of dimension $N\times Kn_t$. The $(i,(k-1)n_t+j)$th 
element of $\mathbf{D}_q$ is the $q$th element of the DFT of the vector 
${[H_{i,(k-1)n_t+j}^{(0)}\,H_{i,(k-1)n_t+j}^{(1)}\,\cdots\,H_{i,(k-1)n_t+j}^{(L-1)}\,0\,\cdots\,0]}^{T}$. 

\vspace{1.5mm}
Performing DFT operation on the received vector $\vy'$ at the receiver,
we get 
\begin{equation} 
{\vz}' \ = \ (\mathbf{F} \otimes \mi_{N}){\vy}' = 
(\mathbf{F} \otimes \mi_{N}) \mh' {\vx'} + \vw',
\end{equation}
where $\vw'=(\mathbf{F} \otimes \mi_{N}) \vn'$.
Further, ${\vz}'$ can be written as
\begin{eqnarray} 
{\vz}' & \hspace{-1mm} = & \hspace{-1mm} \mathbf{D}(\mathbf{F} \otimes \mi_{Kn_t}) {\vx}' + \vw' \nonumber \\
&\hspace{-1mm} = & \hspace{-1mm} \bar{\mh} {\vx'}+\vw', 
\label{cpsc}
\end{eqnarray}
where $\bar{\mh} = \mathbf{D}(\mathbf{F} \otimes \mi_{Kn_t})$ is the 
equivalent channel. Now, detection can be performed on the system 
model in (\ref{cpsc}) using the MP-GSM and CHEMP-GSM algorithms 
presented in Section \ref{sec4}.

\begin{figure}
\hspace{-4mm}
\vspace{-2mm}
\includegraphics[width=3.75in,height=2.75in]{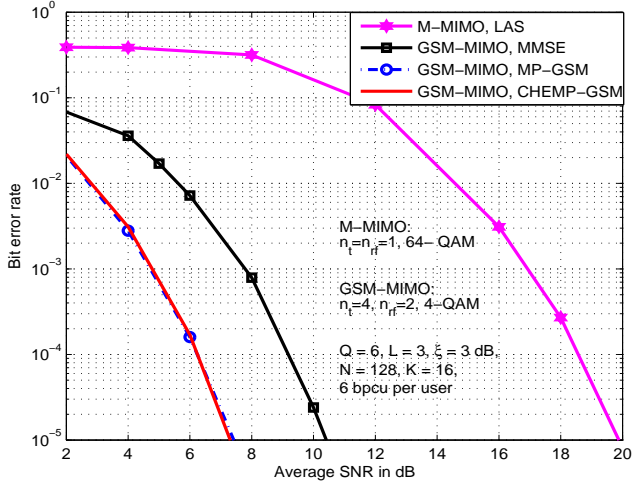}
\vspace{-7mm}
\caption{BER performance of multiuser GSM-MIMO with ($n_t=4$, $n_{rf}=2$, 
4-QAM) and M-MIMO with ($n_t=1$, $n_{rf}=1$, 64-QAM) for $K=16$, $N=128$, 
6 bpcu per user, frequency selective fading with $L=3$ and $\xi=3$ dB, 
CPSC transmission with $Q=6$, and perfect CSI.}
\label{cpsc1} 
\vspace{-4mm}
\end{figure}

\begin{figure}
\hspace{-4mm}
\vspace{-2mm}
\includegraphics[width=3.75in,height=2.75in]{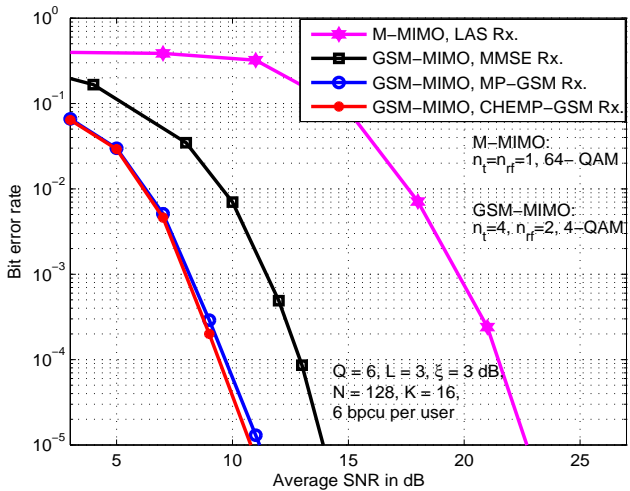}
\vspace{-7mm}
\caption{BER performance of multiuser GSM-MIMO with ($n_t=4$, $n_{rf}=2$, 
4-QAM) and M-MIMO with ($n_t=1$, $n_{rf}=1$, 64-QAM) for $K=16$, $N=128$, 
6 bpcu per user, frequency selective fading with $L=3, \xi=3$ dB), 
CPSC transmission with ($Q=6, L=3, \xi=3$ dB), estimated CSI.}
\label{cpsc3} 
\vspace{2mm}
\end{figure}

{\em Performance with perfect CSI}: 
We evaluated the performance of multiuser GSM-MIMO CPSC systems in frequency 
selective channel with $L=3, Q=6,\xi=3$ dB, $N=128$, and perfect CSI. MMSE, 
MP-GSM and CHEMP-GSM algorithms are used for GSM-MIMO signal detection. 
Figure \ref{cpsc1} shows the performance comparison between GSM-MIMO with 
($n_t=4$, $n_{rf}=2$, 4-QAM) and M-MIMO with ($n_t=1$, $n_{rf}=1$, 64-QAM), 
both having 6 bpcu per user and $K=16$. LAS algorithm in \cite{las} is used 
for M-MIMO signal detection. From Fig. \ref{cpsc1}, we observe that GSM-MIMO 
outperforms M-MIMO by about 12 dB at a BER of $10^{-4}$. Also, MP-GSM and 
CHEMP-GSM detectors outperform MMSE detector by about 3 dB at a BER of 
$10^{-4}$.

{\em Performance with estimated CSI}: 
Figure \ref{cpsc3} shows the performance of GSM-MIMO CPSC systems with
estimated CSI for the same system and channel parameters in Fig. \ref{cpsc1}. 
MMSE receiver (MMSE detector with MMSE channel estimator), MP-GSM receiver 
(MP-GSM detector with MMSE channel estimate, and CHEMP-GSM receiver (CHEMP-GSM
detector with the proposed estimate of ${\bf J}$) are used for GSM-MIMO.
LAS receiver (LAS detection with MMSE channel estimator) is used for
M-MIMO. We observe that GSM-MIMO CPSC system performs better than M-MIMO 
CPSC system by about 11 dB at a BER of $10^{-4}$. 
                
\section{Conclusions}
\label{sec6}
We investigated generalized spatial modulation (GSM) for multiuser 
communication on the uplink in large-scale MIMO systems. We derived 
an analytical upper bound on the average bit error probability in 
multiuser GSM-MIMO systems with ML detection. The bound was shown to 
be tight at moderate-to-high SNRs. Numerical results showed that, for 
the same spectral efficiency, multiuser GSM-MIMO can outperform 
conventional multiuser MIMO by several dBs. 
We also proposed low-complexity algorithms for multiuser GSM-MIMO 
signal detection and channel estimation at the BS receiver 
based on message passing.
The performance of these proposed algorithms in large-scale GSM-MIMO 
systems with tens of users and hundreds of BS antennas showed that 
multiuser GSM-MIMO can outperform conventional multiuser MIMO.
The SNR advantage of GSM-MIMO over conventional MIMO is attributed to the 
following reasons: $(i)$ because of the spatial index bits, to achieve
the same spectral efficiency, GSM-MIMO can use a lower-order QAM 
alphabet compared to that in conventional MIMO, and $(ii)$ to achieve
same spectral efficiency and QAM size, conventional MIMO will need 
more spatial streams per user which results in increased spatial 
interference. This performance advantage along with low RF hardware 
complexity makes large-scale multiuser GSM-MIMO very attractive. 
We further note that the SM concept has 
recently been validated with the aid of experimental activities in 
indoors and outdoors \cite{imp1},\cite{imp2}. These practical 
advancements in SM and the performance advantage in GSM-MIMO suggest 
that large-scale multiuser GSM-MIMO is an attractive technology for 
future wireless systems like 5G.

\vspace{-0mm}
\bibliographystyle{ieeetr}

\end{document}